\documentclass[
reprint,amsmath,amssymb,aps,
prb
]{revtex4-1}

\usepackage{graphicx}
\usepackage{dcolumn}
\usepackage{bm}
\usepackage{amsmath,amsfonts}
\usepackage{mathtools} 
\usepackage{xcolor,url}
\usepackage{placeins}

\newcommand{\JOD}{\textcolor{black}}
\definecolor{NavyBlue}{HTML}{006EB8} 
\newcommand{\RV}{\textcolor{black}}
\newcommand{\RW}{\textcolor{black}}
\usepackage{color,soul}
\usepackage{comment}

\begin{document}
\title{Universal abundance fluctuations across microbial communities, tropical forests, and urban populations}
\author{\vskip -5pt
Ashish B. George$^{1 \ast}$ and James O'Dwyer$^{1 \ast}$ }
\affiliation{\normalsize{$^1$Department of Plant Biology and Carl R. Woese Institute for Genomic Biology, University of Illinois at Urbana-Champaign, Urbana, IL 61801, USA.}}
\author{\vskip -10pt \normalsize{$^\ast$Correspondence to: ashish.b.george@gmail.com, jodwyer@illinois.edu}}
\date{\today}

\begin{abstract}
\noindent The growth of complex populations, such as microbial communities, forests, and cities, occurs over vastly different spatial and temporal scales. Although research in different fields has developed detailed, system-specific models to understand each individual system, a unified analysis of different complex populations is lacking; such an analysis could deepen our understanding of each system and facilitate cross-pollination of tools and insights across fields. Here, for the \JOD{first time we use a shared framework to analyze time-series data of the human gut microbiome, tropical forest, and urban employment}. We demonstrate that a single, three-parameter model of stochastic population dynamics can reproduce the empirical distributions of population abundances and fluctuations in all three data sets. The three parameters characterizing a species measure its mean abundance, deterministic stability, and stochasticity. Our analysis reveals that, despite the vast differences in scale, all three systems occupy a similar region of parameter space when time is measured in generations. In other words, although the fluctuations observed in these systems may appear different, this difference is primarily due to the different \JOD{physical timescales} associated with each system. Further, we show that the distribution of temporal abundance fluctuations is described by just two parameters and derive a two-parameter functional form for abundance fluctuations to improve risk estimation and forecasting. 
\end{abstract}

\maketitle

\section*{Introduction}
\vspace{-10pt}
The dynamics of complex populations is studied in fields ranging from microbiology to economics. These studies have culminated in theoretical and computational models, with various levels of system-specific detail, that have made progress towards explaining system behavior and developing conceptual understanding. This includes the consumer-resource and Lotka-Volterra models of microbial communities; economic, econometric, and physics-based models of urban dynamics; and ecological models of forests based in niche and neutral theory~\cite{macarthur_species_1970,goldford_emergent_2018, hubbell_unified_2001, simini_universal_2012, gabaix_zipfs_1999,acs_measures_1999}. Quantitative analysis, inspired by these models, has also unearthed statistical patterns in the data, hinting at simple emergent behavior of each population~\cite{ji_macroecological_2020, grilli_macroecological_2020,bettencourt_professional_2015,hong_universal_2020, lobo_urban_2020}.

What is lacking, however, is an investigation of emergent behavior using a common framework across different populations and an analysis of the temporal fluctuations in these populations. Classic models in statistical physics, such as diffusion and \JOD{the} Ornstein-Uhlenbeck process, have successfully described the behavior of diverse physical systems~\cite{uhlenbeck_theory_1930,stein_stock_1991,vasicek_equilibrium_1977}. Their success demonstrates that some emergent properties are determined by only a few key underlying details of the dynamics. Efforts at applying this philosophy in other fields have sometimes succeeded~\cite{simini_universal_2012,verbavatz_growth_2020, castellano_statistical_2009,bettencourt_interpretation_2020, mantegna_introduction_2007}, but are often hindered by the lack of data and the incorporation of many system-specific details that make models difficult to analyze. In this paper, we undertake a statistical physics-inspired investigation of three different complex populations: microbial communities, tropical forests, and human cities, through a common framework. Our analysis will aim to uncover the key underlying similarities and differences between the populations; this will not only deepen our understanding of each system, but also facilitate the \JOD{interconnection} of tools and techniques between research fields.

All three complex populations we consider fluctuate in time. Large fluctuations in these populations are associated with catastrophic events such as disease, economic or ecological collapse~\cite{vandeputte_temporal_2021,clark_employment_1998, dai_generic_2012, holling_resilience_1973}; hence understanding these fluctuations is crucial for risk assessment, quantitative biological methods, and forecasting~\cite{west_bayesian_1997, rosenberg_decline_2019,cao_inferring_2017, vandeputte_temporal_2021,holmes_modern_2019,maynard_predicting_2020}. Yet, many models of these populations study the equilibrium and steady-state properties such as the average population abundance
~\cite{holling_resilience_1973,macarthur_species_1970,goldford_emergent_2018,george_ecological_2023, samuels_divergent_1997,fujita_urban_1989, henderson_sizes_1974}. This restriction is in part because 1) analyzing dynamical properties of complex models is harder than analyzing their steady-state behavior and 2) temporal data required to fit and validate complex models has been lacking. We will tackle these two challenges by 1) adopting a minimal model which makes analysis of fluctuations feasible and 2) \JOD{using} data from three separate systems to fit and validate the model. Generating a reliable null model for population fluctuations that enables improved risk estimation is therefore a major goal of the study.

Fitting models using time series analysis has a precedent in each of the fields we consider here. For example, multiple studies have analyzed the possibility that simple models (making a range of assumptions) can reproduce empirical patterns in time-series data of gut microbiome communities~\cite{descheemaeker2020stochastic,grilli_macroecological_2020,ho2022competition}, tropical forests~\cite{azaele_dynamical_2006,Chisholm2014b,kalyuzhny2015neutral,fung2016reproducing}, and urban populations~\cite{verbavatz_growth_2020,simini_universal_2012} (see SI Sec.4). \RW{Despite this history of previous analyses, we do not know of any comparison made across these data types using the same model, hence putting all three data sets on the same footing and amenable to direct comparison.} We'll thus go beyond earlier individual studies by analyzing both time-series and temporal snapshots of data simultaneously, using the same model applied across all three different systems, and introducing novel categorizations of each set of population data into distinct variant types motivated by domain-specific knowledge.

While these three systems span spatial scales from a single body to an entire country, temporal scales from days to years, and are studied in separate research fields, we harness the similarities in structure of the data to identify the same emergent features in all three systems. We compare the observed features to predictions from a model of stochastic population dynamics with just three population-specific parameters. This simplicity allows us to make analytical predictions for emergent features that we can fit and validate using the limited data. Remarkably, the model is able to capture most of the observed variation, despite its simplicity.

\section*{Model and Data}

To develop a unified understanding of complex population dynamics across spatio-temporal scales, we analyze time-series population data from three disparate systems: microbial communities in the human gut, trees in a tropical forest, and employment in U.S. cities (see Fig.1A). Traditionally studied by different fields, we interrogate the data using a shared framework by taking advantage of the similarities in the structure of the data. Each data set contains the relative abundance of population categories within each community at a sequence of time points (see Fig.1B).

\begin{figure*}
\centering
\includegraphics{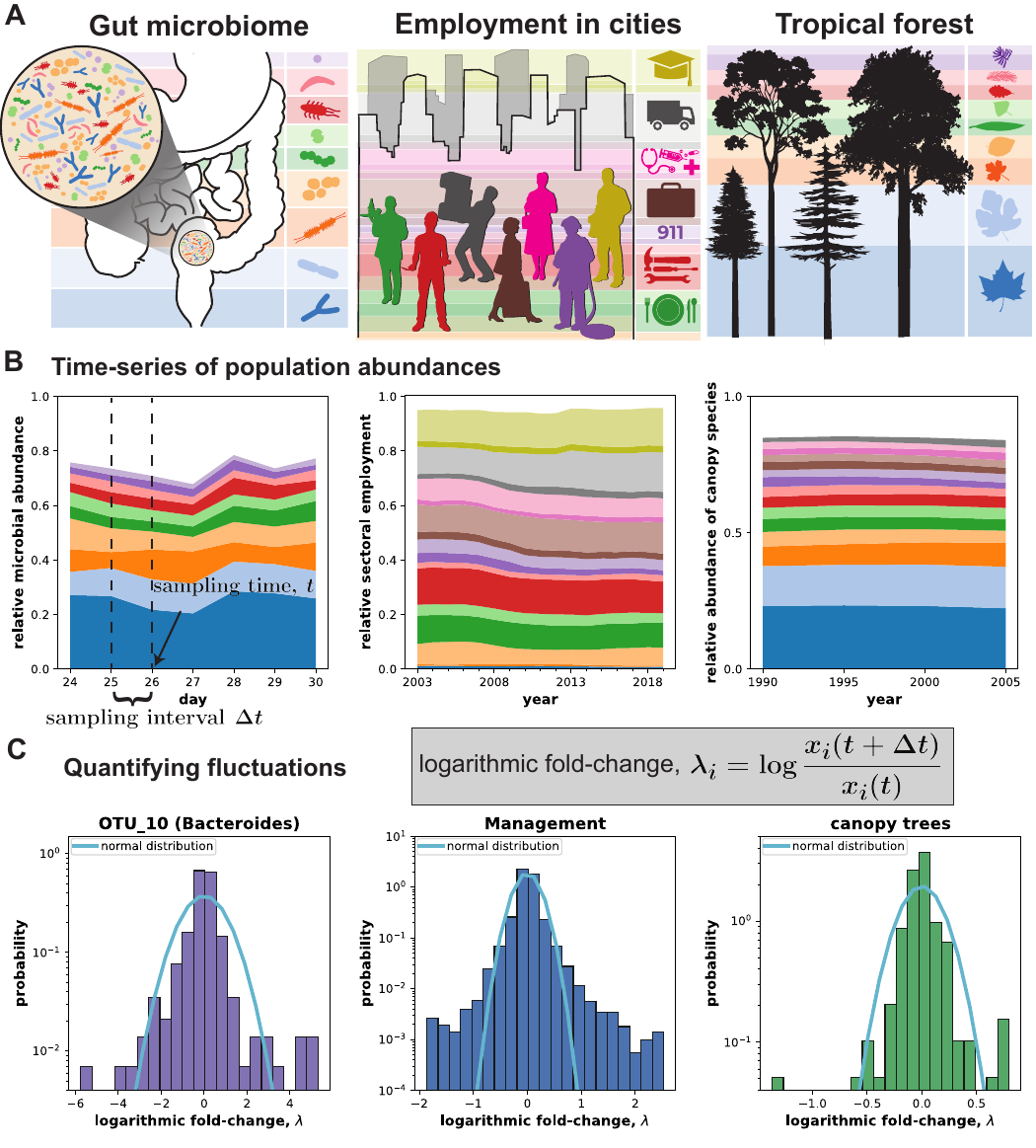}
\caption{\textbf{Time series data from three complex populations spanning spatio-temporal scales.} \textbf{A)} We analyze time-series population data from three complex populations: the human gut microbiome, employment in U.S. cities, and trees in a tropical forest. Each complex population is composed of sub-populations, i.e., microbial species in the microbiome, employment within different economic sectors in cities, and tree species in four trait-based clusters in a tropical forest. \textbf{B)} Abundances of the sub-populations (species/sectors), sampled at periodic intervals, continue to fluctuate in time. Data shows microbiome abundance over a week, employment over 17 years for San Diego metropolitan area, and abundance of canopy tree species. \textbf{C)} The abundance fluctuations are measured using the logarithmic fold-change, $ \lambda_i= \log \frac{x_i(t+\Delta t)}{x_i(t)}$, where $x_i$ is the abundance of microbial species, an economic sector across cities, or tree species in a height cluster. The distribution of logarithmic fold change (LFD) for one microbial species, the management sector across all cities, and all canopy tree species is shown. The fit by the normal distribution to the LFD is shown by the cyan line. The fit is unable to capture the large fluctuations in the tails, central peak of the distribution, or both, illustrating that fluctuations in these complex populations are not normally distributed. Only the 10 (15) most abundant microbial (tree) species are shown in panel B for clarity. 
}
\label{fig:cartoon}
\end{figure*}

More specifically, the gut microbiome dataset records the relative abundances of microbial species in the human gut sampled at daily intervals for almost a year~\cite{caporaso_moving_2011}. \RV{A microbial species or Operational Taxonomic Unit (OTU) was defined based on genomic similarity~\cite{caporaso_moving_2011}.} The data from the Barro Colorado Island (BCI) forest records the number of tree species within a 50 hectare plot on the island, sampled at 5-year intervals for two decades~\cite{condit_complete_2019}. We group trees based on clustering along trait axes into four clusters. \JOD{There is a long history of grouping species by their maximum height~\cite{richards1952tropical}, based on the idea that species with access to similar levels, variability, or horizontal uniformity of light are more likely to compete strongly with each other~\cite{terborgh1985vertical}. Precisely which species to assign to which height cluster has since been put on a more quantitative footing~\cite{dandrea_counting_2020}, with four distinct height clusters (shrubs, understory treelets, midstory trees, and canopy trees) identified on Barro Colorado Island}. The employment dataset \RV{(\url{https://www.bls.gov/cew/})} records the number of people employed in different economic sectors, classified according to North American Industry Classification System (NAICS), in 383 US cities (Metropolitan Statistical Areas). The data is sampled monthly for 17 years.

We analyze the relative \JOD{abundances in all three systems}. Relative employment in a sector (employment in \JOD{the} sector in a \JOD{given} city, divided by total employment in the city) removes the effect of \RV{large variation in total population sizes of cities~\cite{gabaix_zipfs_1999}. Similarly, we analyze the relative abundance of a tree species within a height cluster. Using relative abundances in both these systems allows us to treat data from all three systems on the same footing, normalizes out any temporal changes in the total population sizes stemming from overall population growth or decline (SI Fig.S10), and eliminates the effects of the large differences in city sizes. \JOD{Our Methods section gives} a mathematical definition of relative abundance in each system and SI Sec.8 and Figs.S10-S12 provide our analysis of absolute abundance data. Although working in terms of relative abundances introduces a constraint (that all relative abundances sum to equal one), for diverse communities this constraint has a relatively insignificant effect. We might also expect departures from our model for very high relative abundances, given that this constraint will tend to change the fluctuation properties for such variants as they approach the size of the entire system. But in practice, this is a small effect; it is rare in these data for any relative abundance to even approach $0.5$. For the forest data, we show that our results are robust to the choice of relative vs absolute abundances by analyzing absolute abundances in SI Sec.8 and Figs.S11,S12. }

In contrast to the static predictions from most deterministic models, the data in Fig.1B shows that population abundances continue to fluctuate in time. Furthermore, the strength of these fluctuations differs between the three systems with the largest fluctuations observed in the microbiome. To quantify the strength of fluctuations, we measured the logarithmic fold-change in abundance over a time interval $\Delta t$, defined as
\begin{equation}
\lambda_i = \log \frac{x_i(t+\Delta t )}{x_i(t)},
\end{equation}
where $x_i(t)$ denotes the (relative) abundance of a species/sector $i$ at time $t$. The empirical distribution of logarithmic fold-change or the Logarithmic Fold-change Distribution (LFD) is shown for a single species/sector from each system in Fig.1C. The comparison with the fit of the logarithmic fold-change by the normal distribution (equivalent to fitting fold-change by the lognormal distribution) illustrates that fluctuations cannot be understood as an outcome of environmental noise \JOD{without any additional structure or mechanism}~(see Methods). Further, the fit by the normal distribution indicates that fluctuations much larger than expected from a normal distribution may occur in some of these complex populations. Understanding the distribution of abundance fluctuations is required to quantify the likelihood these large fluctuations which have a major impact on risk-estimates and time-series forecasting.

We now develop a simple model that is capable of predicting statistical features of all three data-sets, including the LFD. To achieve our goal of describing all three data sets, we keep the model both as generally-applicable and simply-formulated as possible. The model assumes that the abundance of a species/sector $i$, $x_i$ fluctuates around an equilibrium value, $x^*_i$, determined by the metabolic, ecological, or economic niche the species/sector occupies. We do not explore the system-specific mechanisms (resource competition, metabolic/economic interactions, etc.) that determine the particular equilibrium value~\cite{macarthur_species_1970,goldford_emergent_2018, horvath_sectoral_2000,fujita_urban_1989, mangan_negative_2010, sardans_empirical_2021}. Fluctuations then occur due to the stochastic processes governing population growth and decline. Deviations from the equilibrium value result in a linear restoring force, described by $(x^*_i-x_i)/\tau_i$, where $\tau_i$ is the timescale of return to equilibrium\RV{, where we neglect additional contributions from species interactions}. Based on these assumptions, we call the model the Stochastic Linear-Response Model~(SLRM) of population dynamics.

Assuming that population growth and decline occur in proportion to the abundance, we can write down the stochastic differential equation of the SLRM governing population abundances:
\begin{equation}
\frac{dx_i}{dt} =\frac{x^*_i}{\tau_i}  -  \frac{x_i}{\tau_i}     + \sqrt{ 2 \sigma_i x_i}\, \eta(t),
\label{eq:dxdt}
\end{equation}
where $\sigma_i$ captures the strength of population fluctuations, and $\eta(t)$ is \RV{delta-correlated Gaussian noise or white noise}. The model resembles the classic Ornstein-Uhlenbeck process, used to describe many stochastic quantities in physics and finance~\cite{uhlenbeck_theory_1930,vasicek_equilibrium_1977, donado_brownian_2017}, with one crucial difference: the scaling of the stochastic fluctuations with the square root of $x_i$. This scaling reflects our assumption on the stochastic processes underlying population growth and decline; specifically, the square root scaling arises when growth and decline is a consequence of many independent, random events and the number of such events is proportional to $x_i$, and is commonly used in many stochastic population models~\cite{lande_stochastic_2003}. \RV{We note that this model has been proposed and analytical solutions were obtained previously in the context of singular diffusion processes by Feller~\cite{feller_two_1951}, bond interest rates by Cox and coauthors~\cite{cox_theory_1985}, and in forest ecology (as a birth-death-immigration model) by Azaele et al~\cite{azaele_dynamical_2006}.} \RW{Here, we go beyond these studies by applying a single model across three different systems.  The SLRM has not to our knowledge been applied individually to employment data or microbiome data, and our application to forest ecology uses populations divided into novel, but ecologically-meaningful categories.}


In ecological terms, the SLRM resembles a scenario where species occupy well-separated niches, with equilibrium abundance $x^*_i$ for species $i$. This idea of niche-separation determining model parameters is reflected in our assumptions. Specifically, in the microbiome, each species is described by its own SLRM parameters; in urban employment, each sector is described by its own SLRM parameters (independent of city); in forests, each trait-based cluster is described by its own SLRM parameters (independent of species index within the cluster). This assumption is reflected in our choice of showing the empirical LFD for a single sector across all cities, a single microbial species in the gut, and all tree species within a trait cluster in Fig.1C. While inter-species interactions may occur in each system, incorporating interactions drastically increases the model complexity hindering parameter-inference. Further, species interactions may not be necessary to describe population fluctuations in these systems, as we will show.

The minimal nature of the model allows us to make analytical predictions for two key quantities that characterize its long-term behavior. First, at long times ($t>\tau_i>0$) the distribution of population abundances will converge to a steady-state distribution. The steady-state distribution arises from the balance of stochastic fluctuations that kick the population from its deterministic equilibrium value and the restoring force towards the equilibrium. The analytical form of the distribution is~(\RV{see Refs.~\cite{feller_two_1951,cox_theory_1985,azaele_dynamical_2006} and SI}) 
\begin{equation}
P_{\mathrm{ss}, i}(x) = \frac{(\sigma_i\tau_i)^{-x^*_i/\sigma_i \tau_i} }{\Gamma(x^*_i/\sigma_i \tau_i)} x^{\frac{x^*_i}{\sigma_i \tau_i}-1} e^{-\frac{x}{\sigma_i\tau_i}},
\label{eq:p_ss}
\end{equation}
which is a Gamma distribution defined by two parameters: $\sigma_i \tau_i$ and the Equilibrium to Noise Ratio (ENR), $x_i^*/\sigma_i \tau_i$, and $\Gamma()$ is the gamma function. $\sigma_i \tau_i$ quantifies the effective strength of the noise: i.e. the fluctuations over the equilibration time period. The ENR, $x_i^*/\sigma_i \tau_i$, quantifies the relative strength of deterministic restoring force to stochastic fluctuations. 

Second, we can analytically derive how abundances \RV{at} steady-state fluctuate in time. Specifically, we quantify the temporal fluctuations by measuring the logarithmic fold-change of population abundances between two time points $\Delta t$ time units apart, denoted as $ \lambda_i= \log \frac{x_i(t+\Delta t)}{x_i(t)}$. \RV{The distribution of logarithmic fold-changes $\lambda_i$ or Logarithmic Fold-change Distribution (LFD), in steady-state was derived in earlier studies~(see Refs.~\cite{cox_theory_1985,azaele_dynamical_2006} and SI})
\begin{multline}
P_{\mathrm{fluc},i}(\lambda, \Delta t ) = \frac{\Gamma\left(\frac{x^*_i}{\sigma_i \tau_i} +\frac{1}{2} \right)}{ 2\sqrt{\pi}  \Gamma\left(\frac{x^*_i}{\sigma_i \tau_i}  \right)} \left(e^{\frac{\Delta t}{\tau_i} } -1\right)^{\frac{x^*_i}{\sigma_i \tau_i}} e^{\frac{\Delta t}{2 \tau_i} }  \\
\cosh{\frac{\lambda}{2}}  \left( \frac{1 }{ e^{\frac{\Delta t}{\tau_i}} (\cosh{\frac{\lambda}{2}})^2  -1 }  \right)^{\frac{x^*_i}{\sigma_i \tau_i} +\frac{1}{2}}.
\label{eq:p_fluc}
\end{multline}
Despite its imposing appearance, the equation depends only on two parameters: the ENR, $x_i^*/\sigma_i \tau_i$, and the equilibration timescale, $\tau_i$, which scales the time interval $\Delta t$. This distribution quantifies the likelihood of large abundance fluctuations in the model. In particular, the tails of the distribution at large $\lambda$ decay exponentially at a rate proportional to the ENR, i.e, $ \sim e ^{-\frac{x^*_i}{\sigma_i \tau_i} |\lambda|}$, which ensures the moments of the distribution are well-defined. Note that this exponential decay for the logarithmic fold-change at large $\lambda$ corresponds to a power-law decay for the fold-change, $\rho= \frac{x_i(t+\Delta t)}{x_i(t)}$, at large $\rho$ with exponent $-\frac{x^*_i}{\sigma_i \tau_i}-1$.

\section*{Results}

\subsection*{Stochastic Linear Response Model~(SLRM) reproduces empirical abundance fluctuations across populations}

\begin{figure*}
\includegraphics[width=\textwidth]{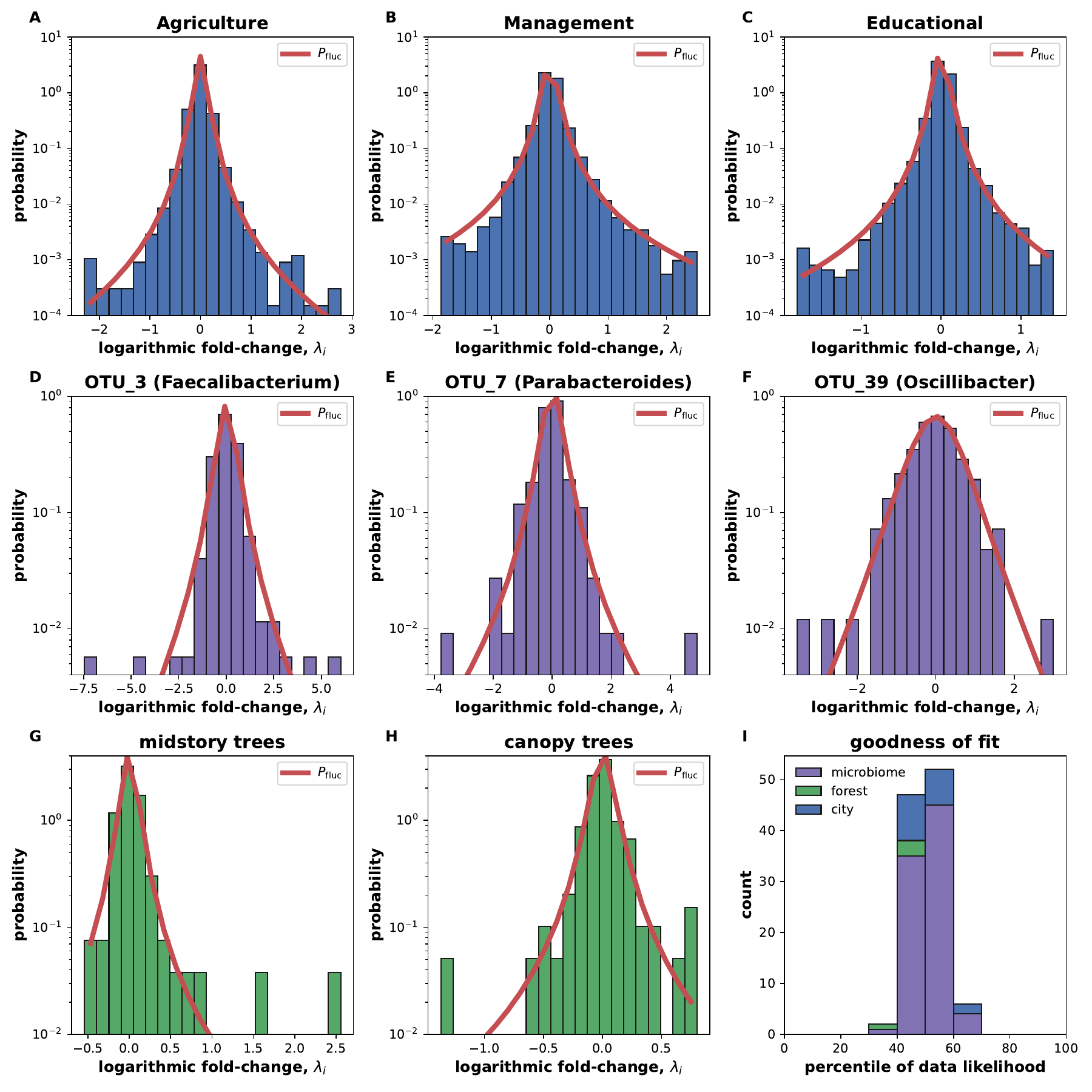}
\caption{\textbf{Universal distribution of fluctuations across complex populations.} The histogram of the empirical Logarithmic Fold-change Distribution (LFD), i.e., the distribution of $ \lambda_i= \log \frac{x_i(t+\Delta t)}{x_i(t)}$, in each system is plotted and fit with the model prediction $P_{\mathrm{fluc}}$ from Eq.~\eqref{eq:p_fluc} (red line). The empirical distribution and fits are shown for:
\textbf{(A-C)} employment in 3 different sectors aggregated across U.S. cities, \textbf{(D-F)} abundances of 3 microbial species in the human gut, and \textbf{(G,H)} abundances of tree species within two height clusters in the BCI forest. \textbf{I)} The percentile score, which quantifies the goodness of fit, shows that the likelihood of observing the data is similar to the likelihood of a random sample from the fitted distribution (of the same size as the data). Specifically, the percentile score quantifies the percentage of random samples with a higher likelihood than the observed data. A percentile score $>95\%$ indicates a poor fit; all fits had a percentile score $<95\%$, as shown in the stacked histogram. The fold-change in abundance was calculated for time intervals of 1 year, 1 day, and 5 years for the city, microbiome, and forest data.}
\label{fig:growth_dists}
\end{figure*}

In line with our goal of characterizing abundance fluctuations in complex populations, we plot the empirical Logarithmic Fold-change Distribution (LFD) in the three systems in Fig.~\ref{fig:growth_dists}. Fig.~\ref{fig:growth_dists}A,B,C plots the LFD of urban employment, aggregated across all US cities, for three different sectors. Fig.~\ref{fig:growth_dists}D,E,F plots the LFD of 3 different species in the human gut microbiome. Fig.~\ref{fig:growth_dists}G,H plots the LFD of all the tree species belonging to two height cluster in the forest. The fold-change in abundance was calculated for time intervals of 1 year, 1 day, and 5 years for the city, microbiome, and forest data, \JOD{respectively}.

All LFDs were roughly centered around $\lambda_i =0$ (average $\lambda_i$ was 0.001, 0.004, and 0.05 for city, microbiome, and forest), but exhibited a large variation relative to the mean (coefficient of variation $\gg 1$; see \RV{SI Fig.S1}). \RV{This relative dominance of fluctuations over systematic trends \JOD{for relative abundances} not only makes the study of fluctuations easier, but also underscores the need to understand fluctuations better.} The Stochastic Linear Response Model (SLRM), which neglects any systematic trends in the mean, is \JOD{therefore} a good candidate model of population fluctuations.

The red lines in Fig.~\ref{fig:growth_dists} show the fit of the observed LFDs with the SLRM prediction, $P_{\mathrm{fluc}}$ (Eq.~\eqref{eq:p_fluc}). A separate fit is performed for each \JOD{employment sector, microbial species, and forest niche}, allowing the inference of corresponding population parameters. Despite having only two free parameters, the SLRM is able to fit the observed fluctuations across these diverse systems. We compared the SLRM fit to fits by two other candidate distributions using the Akaike Information Criterion~\cite{parzen_information_1998}, as an additional test of the fit. The two candidate distributions (normal distribution and Laplace distribution) were chosen based on their success at fitting empirical fluctuations and alternative models~(see Methods and Ref.~\cite{ji_macroecological_2020,stanley_scaling_1996,lande_stochastic_2003}). The SLRM prediction outperformed the other candidate distributions in the majority of cases (\RV{see SI Fig.S2 and datasets 1,2,3}). \RV{We note that our forest analysis demonstrates the importance of the distinct niches, given that inferred timescales range from 400 to 1700 years, and are significantly different between shrubs and other categories~(see SI Sec. S7 and Fig.S13).}

We then further quantified the fits, independent of comparisons with other models, by testing whether the observed data was likely to have been generated by the model. Specifically, for each fit and corresponding parameter estimates, we compared the likelihood of the observed data to the likelihood of 100 random samples of the same size as the data generated from the fitted distribution. \JOD{This kind of `exact' statistical significance test follows a precedent from testing neutral ecological models and population genetics models~\cite{slatkin1994exact,etienne2007neutralsampling}}. If the likelihood of the observed data was smaller than the 95\% of the samples, we concluded that the data was unlikely to have been generated by the fitted distribution and rejected the fit. The percentile score in Fig.1I quantifies the percentage of random samples with a larger likelihood than the observed data. All fits had a percentile score $<$95\%, meaning that SLRM passes our goodness of fit test in all cases. Thus $P_{\mathrm{fluc}}$ serves as a reasonable null expectation for the distribution of population fluctuations.

Comparing the three systems in Fig.~\ref{fig:growth_dists}, we notice that the scale of fold-change on the x-axis appears larger for the microbiome (Fig.~\ref{fig:growth_dists}D-F) than the two macroscopic systems (Fig.~\ref{fig:growth_dists}A-C,G-H). These differences in the data are reflected in the parameters values of the fit, and we explore this parameter variation in the next section. Importantly, despite these differences in scale and shape of the data, $P_{\mathrm{fluc}}$ provides a good fit to data from all three systems.

\subsection*{Universal fluctuations appear different in micro- and macroscopic systems due to \JOD{different generation times}}

Fitting $P_{\mathrm{fluc}}$ to the empirical LFD provides us with maximum likelihood estimates for the equilibration timescale $\tau_i$ and the ENR $x^*_i/\sigma_i \tau_i$. We compare the dimensionless versions of these inferred parameters across the three systems. Specifically, we compare the ratio of the equilibration timescale to time interval of observation $\tau/t_{obs}$ and the ENR across the three systems. 

\begin{figure}
\includegraphics[width=.8\linewidth]{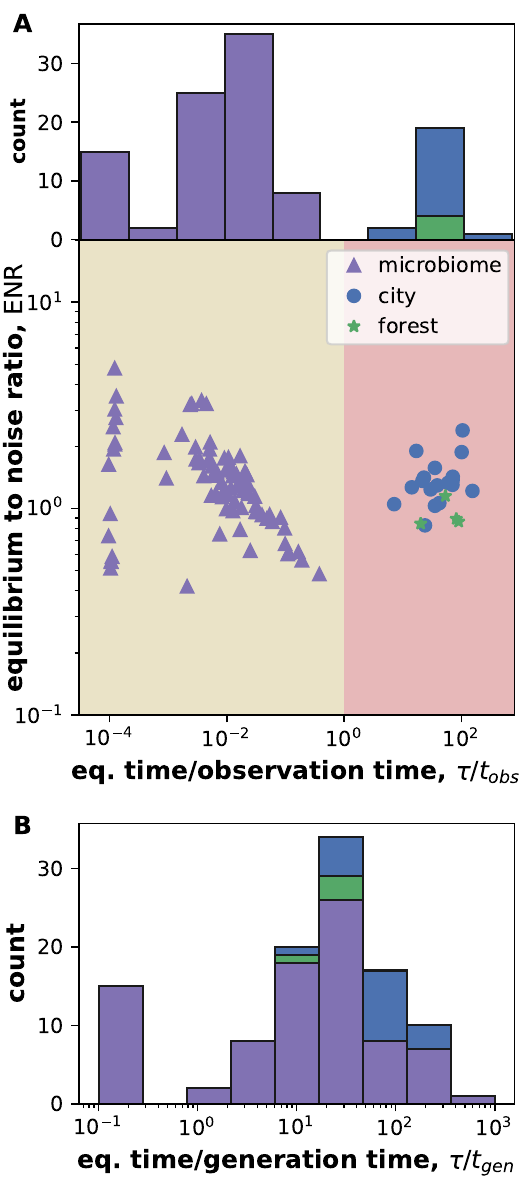}
\caption{\textbf{Timescale of observation differentiates microscopic and macroscopic systems; generation time unifies them.} \textbf{A)} Plot of the dimensionless parameters ENR ($x^*_i/\sigma_i \tau_i$) and $\tau/t_{obs}$ inferred from fitting the empirical LFD. $\tau/t_{obs}$ is the ratio of the equilibration timescale to time duration of observation. All three systems span similar values in ENR, but $\tau/t_{obs}$ differentiates the microscopic and macroscopic systems. Microbiome data has $\tau/t_{obs}<1$, i.e., we observe the microbiome for longer than its equilibration times. The macroscopic systems have $\tau/t_{obs}>1$, i.e., we observe cities and forests for shorter than their equilibration times. \textbf{B)} Measuring time in generations reveals the similarities between systems. The stacked histogram of inferred $\tau/t_{gen}$, the ratio of the equilibration timescale to the generation time, shows that all 3 systems inhabit a similar parameter range, with an equilibration timescale on the order of 10 generations. A small group of microbes appear as outliers with $\tau/t_{gen}<1$; these are the microbes for which the LFD was well-fit by a normal distribution (see \RV{SI Fig.S3}). The estimated generation time, based on Refs.~\cite{gibbons_estimates_1967, lewis_tropical_2004, noauthor_us_2021, hyatt_job--job_2012}, was 4.2 hours, 10 years, and 55 years for microbiome, cities, and forests~(see Methods). \RV{See SI Table S1 for the full name of each sector.}}
\label{fig:logG_params}
\end{figure}

Fig~\ref{fig:logG_params}A plots the ENR and $\tau/t_{obs}$ across the three systems. While the ENR spans a similar range across all three data sets, the micro- and macroscopic systems differ in the inferred values of $\tau/t_{obs}$. For microbial populations, $\tau/t_{obs}<1$, which means that the duration over which we observe the system, about a year, is longer than its equilibration timescale which is around 10 days. In contrast, cities and forests have $\tau/t_{obs}>1$, implying that the duration of observation, two decades, is much shorter than the equilibration timescale. This difference in $\tau/t_{obs}$ is responsible for the difference in shapes of the fitted $P_{\mathrm{fluc}}$ in Fig.~\ref{fig:growth_dists}. \RV{Thus, the duration of observation is an important difference between microscopic (microbiome) and macroscopic systems (cities and forests).}

An alternative way to compare these populations is to use the ratio of the equilibration timescale to the generation time, $\tau/t_{gen}$ (see Fig~\ref{fig:logG_params}B). The generation times in the 3 systems were estimated as $4.2$ hours, $10$ years, and $55$ years for microbes, employment, and forests respectively~(see Methods and Refs.~\cite{gibbons_estimates_1967, lewis_tropical_2004, noauthor_us_2021, hyatt_job--job_2012}). Remarkably, when viewed in terms of generation time, all 3 systems occupy a similar region of parameter space. Hence, fluctuations in the three systems are described by the same distribution over a similar parameter range when time is measured in generations, highlighting the similarities in the emergent behavior across the \JOD{population types}. \JOD{We} note that a small subset of microbes in Fig~\ref{fig:logG_params}B appear as outliers, with $\tau/t_{gen} <1$. For these microbes, the empirical LFD is better-fit by a normal distribution than the model prediction $P_{\mathrm{fluc}}$~(see \RV{SI Fig.S3}). For the remaining data, the equilibration time scale is on the order of 10 generations for all three systems. Therefore, when viewed in terms of generation time rather than \JOD{physical} time, emergent fluctuations in the three systems are highly similar.

\RV{In addition to examining how the inferred timescale, $\tau$, varied across the three systems, we also examined how $\tau$ varied within each system. Analyzing employment data, we found that the most abundant sectors in cities such as healthcare and retail trade had the longest timescales while the least abundant sectors in cities such as agriculture and mining had the shortest timescales. Quantitatively, we found that the median relative abundance of a sector across cities was correlated with the inferred timescale (Pearson's r =$0.86$, Spearman r =$0.66$, $p<0.05$). We found a similar relationship for microbes, with the most abundant species (belonging to Bacteroides) having the longest timescales (Pearson's r =$0.93$, Spearman r =$0.68$, $p<0.05$) (SI Fig.S5). For forest clusters, we found that shrubs had a significantly shorter timescale than the other three, taller height clusters (see SI Sec.7, Figs.S13). We will return to interpret these correlations shortly.}

\subsection*{Stochastic Linear Response Model reproduces empirical distribution of species abundances}
\begin{figure*}
\includegraphics[width=\textwidth]{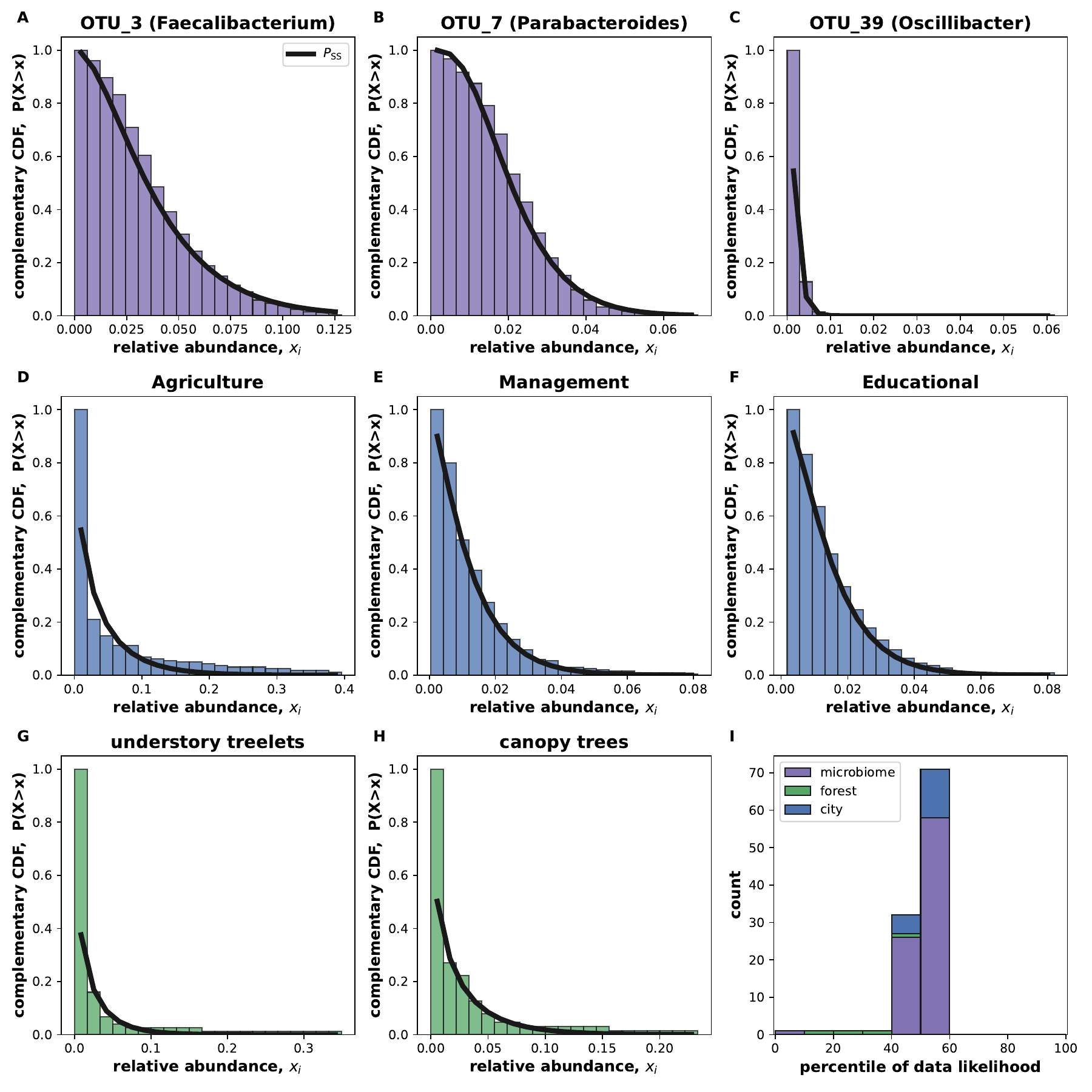}
\caption{\RV{\textbf{The distribution of abundances is well-fit by the SLRM steady-state distribution.} \textbf{A-C)} The distribution of microbial abundances over time for three different species is fit by the two-parameter gamma distribution predicted by the SLRM (black line). We fit the temporal distribution of microbial abundances because we observe the microbiome for far longer than its equilibration time scale. \textbf{D-H)} For cities and forests, we observe the system for intervals shorter than the equilibration time. Hence, we fit the cross-sectional distribution of abundances, i.e, employment in a sector across all cities and abundance of all tree species in a height cluster. Data and fit for the three species, three sectors and two height clusters are shown as the complementary cumulative distribution function $1-CDF(x)$ i.e., the probability that a values i greater than the x-axis $P(X>x)$. \textbf{I)} The percentile score, which quantifies the goodness of fit, shows that the likelihood of observing the data from the fitted distribution is similar to the likelihood of a random sample of the same size as the data. All fits had a percentile score $<95\%$, as seen from the stacked histogram. Two additional goodness of fit tests were also performed and the majority of the species pass both tests (see SI Fig.S6)}}
\label{fig:SSD}
\end{figure*}

From Fig.~\ref{fig:logG_params}, we see that the observation duration for the microbiome is longer than its equilibration time scale ($\tau/t_{obs}<1$). Hence the temporal trajectory of abundances of each microbial species should trace out the corresponding steady-state distribution, which is a gamma distribution described by two parameters (\eqref{eq:p_ss}). We plot the distribution of relative of microbial species and fit it with the two-parameter gamma distribution (black line), via maximum likelihood in Fig.\ref{fig:SSD}A,B,C.

For city and forest data, the observation timescale is longer than the equilibration timescale ($\tau /t_{obs} >1$), and so the temporal trajectory of abundances will not converge to the steady-state distribution. Instead, we plot the cross-sectional distribution of abundances, i.e., the relative abundance of a sector in a city aggregated across all cities and the abundances of tree species within a height cluster, at a random time point~(Fig.~\ref{fig:SSD}D-H). The fit with the gamma distribution is the black line. 

\begin{figure}
\includegraphics[width=\linewidth]{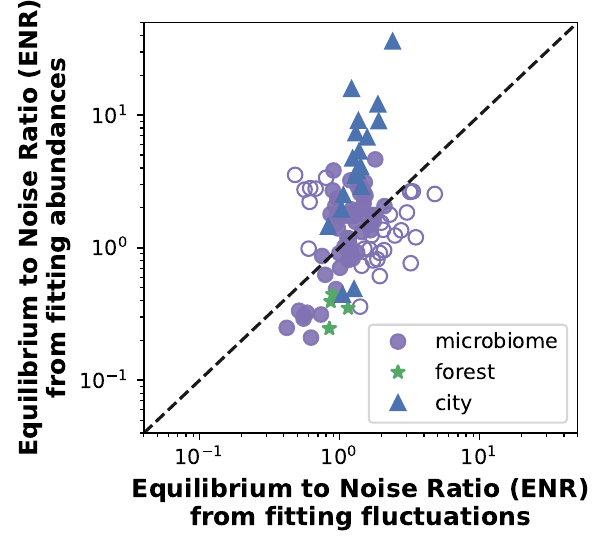}
\caption{\textbf{The consistency of inferred parameters from independent fits of the distributions of fluctuation and abundance.} Independent estimates of the Equilibrium to Noise Ratio (ENR$= x^*_i/\sigma_i \tau_i$) can be inferred from fitting the empirical Logarithmic Fold-change Distribution (LFD) or ftting the abundance distribution. If a fit of the LFD using the ENR value inferred from fitting the abundance passes the goodness of fit test, or vice-versa, we deem the inferred values of ENR as consistent. Filled points denote consistent estimates and unfilled points denote inconsistent data points. All city sectors and forest species were consistent, and 57 of 85 microbial species were consistent.}
\label{fig:parameter_consistency}
\end{figure} 
To test whether the observed data could have been generated by the model, we repeat the procedure used in Fig.~\ref{fig:growth_dists}. Specifically, we compared the likelihood of the observed data to the likelihood of randomly generated samples from the fitted distribution. All data sets had a likelihood comparable to a random sample, as evidenced by the percentile scores (from 1000 random samples of the same size) shown in Fig.\ref{fig:SSD}I. \RV{Two additional goodness of fit tests were also performed and the majority of species passed both tests (see SI Sec.S5 and Figs. S6, S7)}. Together, these observations indicate that the abundances in these systems are well described by the gamma distribution. This conclusion is further supported by previous research that used a gamma distribution to successfully fit cross-sectional microbial abundances across microbiomes~\cite{grilli_macroecological_2020}. 

The steady-state distribution~(SSD) and logarithmic fold-change distribution~(LFD) predicted by the model (\eqref{eq:p_ss},~\eqref{eq:p_fluc}) share a parameter, the ENR ($x^*_i/\sigma_i \tau_i$). We compare the maximum likelihood estimates of the ENR obtained by the separate fits of the empirical abundance distribution and LFD in Fig.~\ref{fig:parameter_consistency}. The discrepancy in inferred ENR values could arise from limitations of the data or the model.

To check whether the model is able to simultaneously fit both the LFD and abundance distribution, we performed a modified version of our goodness of fit test we used previously. First, using the ENR inferred from fitting the abundance, we compare the likelihood of fitting the LFD by $P_{\mathrm{fluc}}$ to the likelihood of 100 random samples from the fitted distribution of the same size. Then we fit the other way around, i.e., using the ENR inferred from fitting the LFD, we compare the likelihood fitting the abundance by $P_{\mathrm{ss}}$ to the likelihood of 1000 random samples from the fitted distribution of the same size. If the likelihood of the observed data was within 95\% of the likelihood of the samples for either of these comparisons, then we conclude that the inferred parameters were consistent, i.e., the ENR obtained from fitting one distribution is able to provide a reasonable fit of the other distribution. The data points deemed consistent are depicted by filled markers in Fig.~\ref{fig:parameter_consistency}. If the data likelihood was smaller than 95\% in both cases, we term the inferred parameters as inconsistent. The data points deemed inconsistent are shown as unfilled markers in Fig.~\ref{fig:parameter_consistency}.

The ENR estimates for the majority of the data were consistent (Fig.~\ref{fig:parameter_consistency}). A minority of the microbes (28/85) were rejected as being inconsistent. This could partly be due to temporal correlations in the data used to fit the abundance distribution, which we neglected. These correlations exist over timescales of $\tau_i$ and only vanish when $\Delta t \gg \tau_i$. For cities and forests, we used cross-sectional data, which does not suffer from this drawback. All employment data, including the apparently large outliers in Fig.~\ref{fig:parameter_consistency}, were consistent. The consistency of these large outliers was because although the abundance distribution of some sectors were well fit by large ENR values, substantially smaller ENR values also provided a reasonable fit and could not be rejected using the few hundred observations. Overall, the majority of the observed species/sectors in the three systems followed the expected relationship between the predicted distributions for abundances and fluctuations.

\subsection*{Variation of model parameters explains Taylor's law}

\begin{figure}[t!]
\includegraphics[width=.8\linewidth]{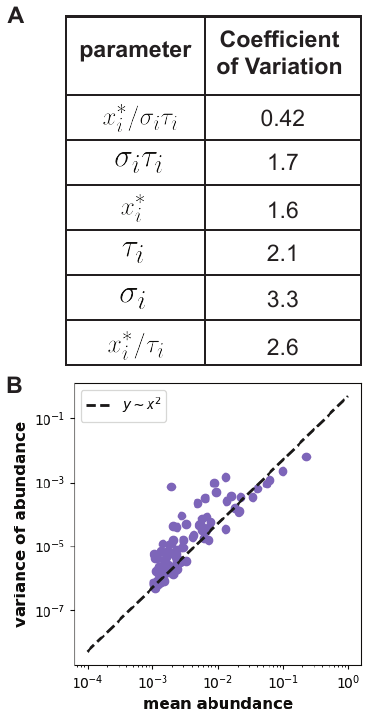}
\caption{\textbf{Power-law scaling of temporal mean and variance of abundance (Taylor's law) in microbiome due to approximate conservation of ENR.} \textbf{A)}The inferred ENR ($x^*_i/\sigma_i \tau_i$) in the microbiome data is relatively constant compared to other parameters, as evidenced by the tabulated coefficient of variation. \textbf{B)} The properties of the gamma distribution (which fits the abundance distribution) define the mean abundance of species $i$, as $x^*_i$, and the ratio of variance to mean squared as the ENR $x^*_i/\sigma_i \tau_i$. Thus, due to the approximate conservation of the ENR, we get the power-law scaling (with exponent 2) of variance with mean referred to as Taylor's law. The geometric mean of the ENR inferred from fitting the LFD and abundance distribution was used when estimating the variation of different parameter combinations.}
\label{fig:taylors_law}
\end{figure}

In addition to providing a simple 2-parameter null model for fluctuations and abundances in complex populations, the SLRM can also help understand other empirical patterns in the data. Prior research in microbiome data has revealed an approximate power-law scaling (with exponent 2) of the variance vs. mean of the abundance, called Taylor's law~\cite{grilli_macroecological_2020, ji_macroecological_2020}~(Fig.~\ref{fig:taylors_law}), which may arise due to various mechanisms~\cite{cohen_random_2015,giometto_sample_2015,fronczak_origins_2010}. Examination of inferred model parameters provides an alternative explanation for this empirical observation. 

To investigate why Taylor's law arises, we compare the variation of the inferred model parameters across microbial species, using the coefficient of variation~(Fig.~\ref{fig:taylors_law}A). The ENR ($x^*_i/\sigma_i \tau_i$) has a substantially lower coefficient of variation than the other parameters, and so can be considered to be approximately constant. \JOD{For} the Gamma distribution, the mean abundance of a species $i$ is $x^*_i$ and the ratio of variance to mean squared is the ENR, $x^*_i/\sigma_i \tau_i$. In microbiome data, the ENR is constant, and so the ratio of variance to mean squared remains fixed while the mean abundance varies, which leads to the observation of an approximate power-law scaling of the variance with mean. Hence, the approximate constancy of the ENR provides an alternative explanation for the observed power-law known as Taylor's law in microbiome data. The approximate constancy of ENR implies that the distribution of fluctuations (LFD) of different microbial species will be similar when time intervals are measured in terms of $\tau_i$. \RV{Furthermore, the approximate constancy may explain one of our earlier findings---that relative abundance correlates with inferred timescale (SI Fig.S5). Mathematically, since the ENR does not vary significantly, relative abundance in our model must be proportional to $\sigma_i \tau_i$. Future work may uncover the mechanisms behind why ENR is approximately constant across these systems.}

\subsection*{\RV{Comparing SLRM and a model with environmental noise}}
\RV{The Stochastic Linear Response Model (SLRM) incorporates `square-root' fluctuations, referred to as demographic noise, and commonly used in many population dynamics models~\cite{lande_stochastic_2003}. Demographic noise captures the fluctuations arising from accumulation of small, independent random growth and death events, and mathematically it can be identified by the square-root scaling of the noise term with population size (Eq.~\ref{eq:dxdt}). An alternative form of noise used in population dynamics models is environmental noise~\cite{lande_stochastic_2003}. }\RW{Environmental noise aims to capture fluctuations arising from random fluctuations of the overall growth and death rates of the population, and has been used in the analysis of both microbiome data~\cite{grilli_macroecological_2020} and local forest communities~\cite{Chisholm2014b,kalyuzhny2015neutral,fung2016reproducing}. The latter involve a range of choices of model specification, including the way competition is imposed in the local community, and how dispersal is modeled from regional pool to local patches. This complexity tends to yield models without analytical solutions for LFD and SSD. On the other hand, a relatively simple implementation of environmental stochasticity is the Stochastic Logistic Model (SLM), which has been applied to recapitulate the observed abundance distributions in another of our three data types: microbiome communities~\cite{grilli_macroecological_2020, descheemaeker2020stochastic}. All of these models are characterized by the same linear scaling of the noise term in the population size ~(Eq.~\ref{eq:dxdt_SLM}), and so comparing our model with an environmental noise model provides an initial test of whether environmental stochasticity will inevitably tend to provide a better description of fluctuations than demographic noise alone. Therefore, in this section we compare the SLRM, which incorporates demographic noise, with the SLM.}

\RV{The SLM is also a three-parameter model defined by the following equation for the relative species abundance $x_i$:}

\RV{\begin{equation}
\frac{d x_i}{d t}=\frac{x_i}{\tau^{\prime}_i}\left(1-\frac{x_i}{K_i}\right)+\sqrt{\frac{S_i}{\tau^{\prime}_i}} x_i \eta_i(t),
\label{eq:dxdt_SLM}
\end{equation}
with parameters $K_i$, which describes the carrying capacity of the population, $S_i$, which captures the strength of fluctuations, and $\tau^{\prime}_i$, which sets the timescale of growth and equilibration. $\eta_i$ is delta-correlated Gaussian noise or white noise. }

\RV{The Steady-State abundance Distribution(SSD) of the SLM is also a Gamma distribution~\cite{grilli_macroecological_2020}, like the SLRM:
\begin{equation}
P^{(SLM)}_{\mathrm{ss}, i}(x) = \frac{x^{2S_i^{-1}-2} }{\Gamma(2S_i^{-1}-1)} \left(\frac{2}{K_iS_i} \right)^{2S_i^{-1}-1}  e^{-\frac{2 x}{K_iS_i}}.
\label{eq:p_ssSLM}
\end{equation}
It is parameterized by combinations of the two parameters $K_i$ and $S_i$. However, unlike the SLRM, there is no analytical prediction for the Logarithmic Fold-change Distribution of the SLM.}

\RV{Therefore, to facilitate a direct comparison between the SLRM and SLM, we adopted the following procedure: first, we fixed two of the three parameters in both models by fitting the gamma-distributed SSD predicted by the models to the observed abundance distribution. Then, we simulated the SLM for a range of values of remaining parameter, $\tau^{\prime}$, to obtain a series of predicted LFDs from SLM simulations. We obtained the LFD for the same range of $\tau$ values of the SLRM through the analytical predictions. Finally, we compared the disagreement between the two sets of predicted LFDs and the empirical LFDs in the three systems by computing the Jensen-Shannon Distance between them (see Methods for further details).}

\begin{figure*}
\includegraphics[width=\textwidth]{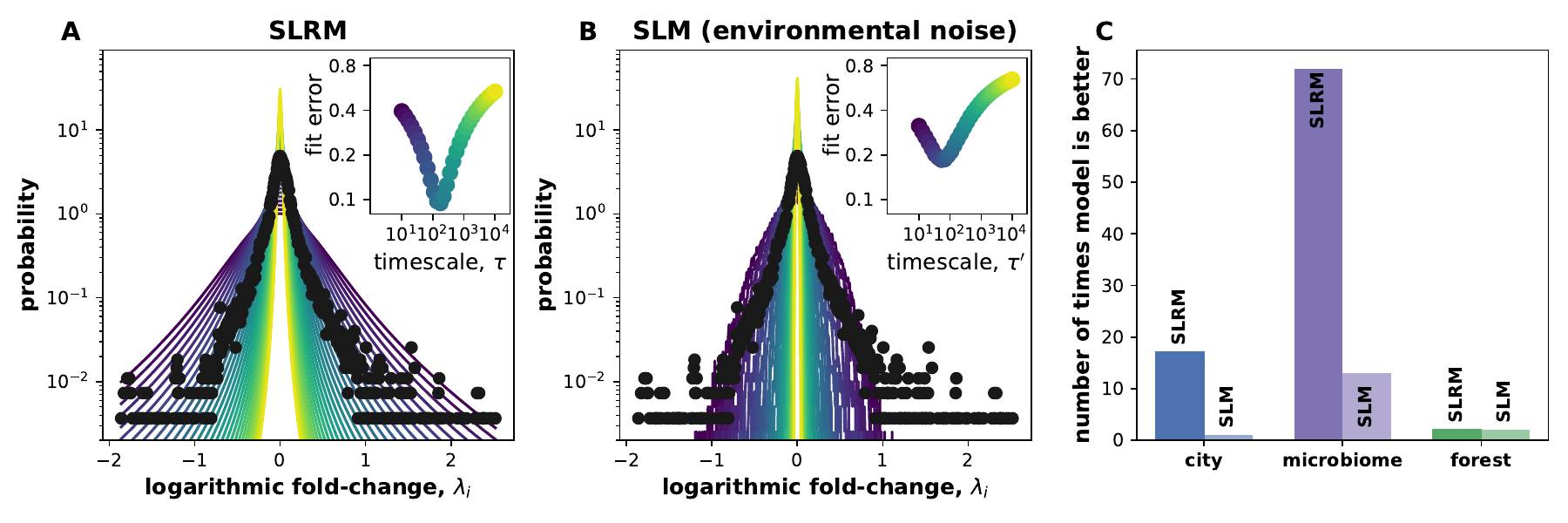}
\caption{\RV{\textbf{Comparing SLRM and a model with environmental noise (SLM).} \textbf{A,B)} We compare the predicted Logarithmic Fold-change Distribution (LFD) of the SLRM, which incorporates demographic noise, and the SLM, which incorporates environmental noise, to the empirical LFD (black points). Colored lines in panels A and B show the LFD of the SLRM and SLM varies for the same range of parameter ($\tau^{\prime}$,  $\tau$) values; empirical data corresponds to employment in the management sector. The insets show the error between the model predictions and observed data, measured using the Jensen-Shannon Distance. A lower value of the JSD indicates a better fit; the SLRM fit this data better than the SLM. \textbf{C)} We repeated this procedure for all sectors, species and clusters to identify the model that provided a better fit (lower JSD) in each case. The SLRM provided a better fit to majority of the data---in 17 out of the 18 employment sectors, 72 out of 85 microbial species, and 2 out of the 4 forest clusters.}}
\label{fig:compare_SLRM_SLM}
\end{figure*}

\RV{In Fig.~\ref{fig:compare_SLRM_SLM}A,B, we illustrate this procedure applied to data on the employment in the management sector in US cities. The distributions predicted by the two models as the timescale parameters ($\tau^{\prime}$,  $\tau$) are varied is shown by the colored lines alongside the observed data (black circles). The disagreement between the model prediction and the observed data is measured using the Jensen-Shannon Distance (JSD) and shown in the insets. The model with a lower JSD better explains the observed data. We repeated this analysis for all microbial species, employment sectors and forest niches, and found that the SLRM provided a better fit in the majority of the cases (Fig.~\ref{fig:compare_SLRM_SLM}C). Specifically, the SLRM had a lower JSD in 17 out of the 18 employment sectors, 72 out of 85 microbial species, and 2 out of the 4 forest clusters. Interestingly, it is the forest data, where a body of evidence exists to demonstrate the importance of environmental noise in explaining other aspects of population fluctuations~\cite{fung_reproducing_2016,kalyuzhny_neutral_2015}, where the SLRM and SLM were most similar in their performance. But outside of these cases, the SLRM provides the better description of the data.}

\section*{Discussion}

Understanding the behavior of diverse complex populations is challenging, but critical for progress in many fields. \JOD{For the first time, we analyze all three of microbiome data, forest data, and employment data on an equal footing. This unified approach to studying the three populations complements} detailed investigations of each specific population. \RW{In the microbiome context, where many studies focus on species interactions~\cite{grilli_higher-order_2017, ho_competition_2021}, our work demonstrates how statistical features of the data can arise from stochastic fluctuation, with minimal interactions. For forest data, our work finds some independent support for prior analysis that groups trees into distinct height niches~\cite{dandrea_counting_2020}. And for employment data, where many studies have focused on detailed econometric models and analyses~\cite{horvath_sectoral_2000, acs_measures_1999, armington_job_2004}, our work provides a novel perspective from the point of view of a simple dynamical model.} Moreover, by showing that the emergent distributions of population abundances and fluctuations in all three populations can be reproduced by a single model, our analysis highlights emergent properties that are independent of system-specific details. When using generations to measure time, all three systems occupy a similar range of parameters, suggesting that fluctuations in cities and forests over decades and centuries would closely resemble microbiome fluctuations. \RW{In broader terms, the fact that our analysis provides a good description across a variety of systems may point to the existence of universal properties in the fluctuations of complex systems.}

\JOD{In quantitative terms}, our analysis provides a simple two-parameter functional form for the distribution of fold-changes in the populations and relates this to the abundance distribution. The predicted distribution of fluctuations $P_{\mathrm{fluc}}$~(\eqref{eq:p_fluc}) is able to fit fluctuations in all three systems despite having only two parameters, \JOD{and moreover we find that when measured in terms of generation time, parameter fits for all three systems collapse into a narrow window of fitted values.} Further, $P_{\mathrm{fluc}}$ arises from a well-defined model and suggests a plausible mechanism; it is not simply chosen from the vast library of statistical distributions historically examined. \JOD{The observation that we find all three data types are described by similar parameter values suggests important, deeper connections that may yet be uncovered in future work}.

\RV{The predicted distribution of fluctuations $P_{\mathrm{fluc}}$ can serve as an important null model of the fluctuations} in complex populations, where understanding the likelihood of large fluctuations is crucial. For employment fluctuations, large fluctuations impact urban planning and economic stability; for forests, large fluctuations impact ecological management strategies; for microbiome, large fluctuations can cause dysbiosis, which affects the health of the host~\cite{vandeputte_temporal_2021,clark_employment_1998,holling_resilience_1973}. A two-parameter null model for fluctuations in complex populations is useful in practical, data-limited settings; \RW{it can estimate the risk of large fluctuations more accurately~(SI Fig.S4) and improve quantitative methods, such as those utilising Bayesian inference from time-series data to classify ecosystem states~\cite{bogart_mitre_2019} and priors for
priors for decision-making and modeling~\cite{robert_bayesian_2007, shafiei_biomico_2015, hwang_prototype_1992}.} \RV{In the SI, we show how $P_{\mathrm{fluc}}$ fits empirical data from the three systems when fluctuations are measured over different time intervals (see SI Sec.S6, Figs. S8,S9).}

The Stochastic Linear Response Model (SLRM) can be understood as the linearization of a more complex nonlinear model around its equilibrium when species interactions are neglected. \RV{This is shown in our SI, where we also demonstrate (in SI Sec.S2) an example where a model with inter-species interactions actually reduces exactly to the SLRM. In general though, this simplification drastically reduces the number of parameters, making parameter inference from available data feasible. Deviations from the model predictions, could indicate the presence of species interactions, which are often modeled by Lotka-Volterra, consumer-resource and other models of higher order species interactions~\cite{goldford_emergent_2018,macarthur_species_1970,grilli_higher-order_2017}, \JOD{autocorrelated noise}, or other mechanisms}. Such models could potentially be parameterized by using specialized methods with additional data~\cite{fisher_identifying_2014,cao_inferring_2017}. \RW{In the SI, we discuss how the SLRM can be extended into a stochastic model incorporating species interactions in a linear regime. Analysis of this extended model could pave the way for novel inference methods that account for the stochastic fluctuations in observational data.}

\JOD{While we compared our model predictions with a range of classic distributions, we also note that environmental noise~\cite{lande_stochastic_2003} has been proposed as an explanation for fluctuations in abundance across different complex systems, \RW{including multiple forest data sets~\cite{Chisholm2014b,kalyuzhny2015neutral,fung2016reproducing} and microbiome data~\cite{grilli_macroecological_2020}}. To capture this mechanism and compare our model to its predictions, we tested the performance of the SLRM to the Stochastic Logistic Model (SLM)~\cite{grilli_macroecological_2020,descheemaeker2020stochastic}, a three-parameter model that combines nonlinear logistic growth with environmental stochasticity. The SLM lacks an analytical solution for the Logarithmic Fold-change Distribution, but through numerical simulations we compared the fits of the SLRM and SLM to the empirical data. We found that the SLRM outperformed the model with enviromental noise in the majority of our data~(Fig. ~\ref{fig:compare_SLRM_SLM})}. \RW{While there are multiple other types of enviromental noise model, for example those that have provided a good description of local forest community fluctuations~\cite{kalyuzhny2015neutral,fung2016reproducing}, this comparison demonstrates that environmental stochasticity does not necessarily provide a better description of fluctuations in complex populations. More general models of environmental stochasticity tend to lack simultaneous analytical solutions for the Logarthmic Fold-change Distributions and Steady-State Abundance distribution, making numerical comparison more challenging.  However, it is certainly possible that, just as with species interactions, more general kinds of noise should form part of the basis for extending our model, and future analysis will likely shed light on this question.}

\JOD{Framing the SLRM as a useful base model} for further research, we note that it can be easily augmented with additional mechanisms, including environmental fluctuations and species interactions, which could be tested with additional data~(see SI). Other modifications could help understand evolving populations. For the timescales examined, we assumed that the equilibrium value $x^*$ remains constant. Over longer timescales, however, the equilibrium value could change due to biological evolution, climate change, or socio-technological revolution. Investigating the model when $x^*$ changes in time could help understand emergent dynamics in complex populations over evolutionary timescales and presents an interesting direction for future research.

To butcher two well-worn phrases, all models are wrong, some are useful, and some are unreasonably effective. We believe the SLRM falls into the latter two categories, and that its surprising effectiveness across such diverse datasets points to something universal about the way complex populations fluctuate. The SLRM also provides valuable two parameter null models for the distributions of abundances and fluctuations in complex populations, which are of particularly utility in data-limited scenarios for forecasting and risk-analysis. The unified analysis of the three \JOD{population types highlights both similarities and differences between the systems, and paves the way for a fruitful exchange of tools, techniques, and interpretations between these very different fields}.

\section*{Methods}
\subsection*{Data processing}

\subsubsection*{City data}
Public domain city data were obtained from Quarterly Census of Employment and Wages from the U.S. Bureau of Labor Statistics (https://www.bls.gov/cew/). The data provides the employment classified into industrial sectors by the North American Industry Classification System (NAICS) at county level in the U.S.~(see SI Table S1). We aggregated data at the county level to 383 Metropolitan Statistical Areas (MSA), which we call cities. MSAs are independent statistical units defined by the Census Bureau that encompass a central city and the geographical areas connected to the city. We obtained the list of counties in each MSA in 2017 from U.S. Census Bureau, County Business Patterns program, and used this to calculate employment at MSA level. This approach allowed us to maintain a consistent definition of MSAs across the entire time-series. The employment data is recorded at monthly intervals from 2003 to 2019. Since many industries, such as agriculture and accommodation, display seasonal trends in employment withing a year, we used a $\Delta t$ of 1 year for calculating the empirical LFD. \RW{The duration of observation, $t_{obs}$, was 17 years for employment data.} We plot and fit data with only non-zero abundance.

For privacy reasons, sectoral employment data at some points are suppressed, and so we remove these points from our analysis. Since this suppression increases at finer levels of NAICS classification, we analyzed sectors classified at the two digit level. This also made our analysis robust to changes in the NAICS classification scheme that impacted the temporal continuity of the data at finer resolution. We used 18 of the 21 NAICS categories at the two-digit level (see SI Table S1). We removed three categories: `81' (other) ,`92' (public administration) and `99' (unclassified). Public administration and government employment was removed because much of this sector is not reported due to governmental regulations. Relative sectoral employment was calculated for each city by dividing sectoral employment in the city with the total employment reported in the city. Note that due to data suppression and removal of three naics categories, the sum of the relative employment in a city over all sectors we analyze need not equal one. We assume that each sectors is described by a set of SLRM parameters that is independent of the city; this provides us with enough data to fit the SLRM predictions.

\subsubsection*{Microbiome data}
Microbiome data from Ref.~\cite{caporaso_moving_2011} was obtained and processed as in Ref.~\cite{ji_macroecological_2020}. We consider only the gut microbiome data for individual M3 since it was substantially longer than other time-series. \RW{The data was collected in time intervals of one day with some gaps in sampling, and hence $\Delta t =1$ day for microbiome data. There were a total of 336 time-points recorded excluding sampling gaps. Since there were sampling gaps, an approximate duration of observation, $t_{obs}$, of 300 days was used for microbiome data.}

To process the data, first, read counts at each time point were normalized to obtain relative abundance. Only prevalent or relatively abundant species were used for our analysis. Specifically, we considered species only if they were present in more than half of the time points and their average abundance was greater than $10^{-3}$. 
85 species that met this criterion. To calculate the empirical LFD, we used abundances that were collected $\Delta t=1$ days apart. We plot and fit data with non-zero relative abundance. Each species has its own set of SLRM parameters.

\subsubsection*{Forest data}
The Barro-Colorado Island forest data was obtained from the Center for Tropical Forest Science website (https://forestgeo.si.edu) ~\cite{condit_complete_2019}. Abundance data collected at 5 year intervals, in years 1990, 1995, 2000, and 2005 was used. Only trees that were alive and had a diameter at breast height $>10$ cm were counted. The trees were grouped into four height clusters: shrubs, understory treelets, midstory trees, and canopy trees based on Ref.~\cite{dandrea_counting_2020}. There were 87, 75, 60, and 63 species in the four height clusters. Relative abundance of a species in a particular height cluster was calculated as the absolute abundance of the trees of the species at that time point divided by the absolute abundance of all tree trees within that specific height cluster. We assume that all species within a height cluster and thus being highly similar in trait values, are described the same set of SLRM parameters; this provides us the data required to fit the data with the model predictions. Different height clusters are fit separately, like different employment sectors. Time interval for calculating LFD, $\Delta t$, was 5 years. \RW{The duration of observation, $t_{obs}$, was 20 years for forest data.}

\subsection*{Fitting and sampling procedures}

The data was fit and parameters were estimated by Maximum Likelihood Estimation from the Scipy package in Python. In addition to the probability distribution $P_{\mathrm{fluc}}$, the corresponding cumulative distribution (\RV{see SI}) was defined to make sampling from the distribution more efficient. 

In the city data set, we had substantially more data points in the each LFD~($\sim 30,000$ points) than for the empirical abundance distribution~($\sim 300$ points). This made performing the goodness of fit test computationally harder for the LFD than the abundance distribution, since the test required generationg samples from the distribution of the same size as the data. Hence for each fit of the abundance distribution, we obtained 1000 samples to compare with the data likelihood, whereas for each fit of the LFD, we obtained only 100 samples.

The coefficient of variation of tabulated in Fig.~\ref{fig:taylors_law}A, used inferred parameters values of ENR, $\sigma_i \tau_i$, and $\tau_i$ to construct the various parameter combinations. $\tau_i$ was estimated from fitting the LFD, $\sigma_i \tau_i$ was estimated from fitting the abundance distribution, and the ENR was the geometric mean of the estimates from the LFD and abundance distribution.

\subsection*{\RV{Definition of relative abundance in each system}}

\RV{For microbiome data, the relative abundance of species $i$ at time $t$, $x_i(t)$ was defined as
\begin{equation}
x_i(t)= \frac{n_i(t)}{\sum_j n_j(t)},
\end{equation}
where $n_i(t)$ is the number of sequence read counts of the species obtained at time $t$. Since the number of read counts is different from the species number, we do not know the absolute abundance of microbes in the data.\\
For the forest data, relative abundance of species $i$ in cluster $c$ at time $t$, $x_i(t)$, is given by
 \begin{equation}
x_i(t)= \frac{n_i(t)}{\sum_j n_j(t)}, \, \mathrm{s.t.}\;  \{i,j\} \in c, 
\label{eq:rel_abu_city_forest}
\end{equation}
where $n_i(t), n_j(t)$ are the counts of species $i,j$ belonging to the same cluster $c$ at time $t$. Hence, species abundance is normalized by the total population within the same height cluster to obtain its relative abundance.\\
Eq.~\ref{eq:rel_abu_city_forest} can also be used to define relative abundance in city data. For city data, $x_i(t)$ is the relative abundance of sector $i$ in city $c$ at time $t$ and $n_i(t), n_j(t)$ is the employment in sectors $i,j$ in the same city $c$ at time $t$. Hence, sectoral employment is normalized by the total employment in the same city to obtain its relative abundance. This allows us to neglect the wide variation in total population sizes across cities~\cite{gabaix_zipfs_1999}.}

\subsection*{Estimating generation times in each system}

The generation times used for scaling the inferred time scale in Fig.~\ref{fig:logG_params} were calculated based on estimates in the literature. Since a direct estimate of generation time of microbes in the human gut is unavailable, we used the measured generation time in mice on a regular diet of 5.7 divisions per day(4.2 hours)~\cite{gibbons_estimates_1967}. For trees, the generation time used was 55.5 years~\cite{lewis_tropical_2004}. For urban employment, we measured generation time as the time required for people to change jobs between industrial sectors. Using the median job duration in US of 5 years~\cite{noauthor_us_2021} and the fact that roughly half of the job changes are between sectors~\cite{hyatt_job--job_2012}, we obtained a generation time of 10 years. The estimated generation times neglect possible variation between species and sector due to data limitations.

\subsection*{\RV{Simulating a model with environmental noise (SLM)}}

\RV{We implemented the Stochastic Logistic Model (SLM) using a temporal finite difference scheme in python. To ensure accurate simulation results, we chose time steps that were sufficiently small. Instances of the noise, $\eta$ were generated by sampling a normal distribution with variance scaling appropriately with the time-step. To avoid species extinction, we imposed a small minimum population size. We simulated the population for longer than $10 \tau^{\prime}$ to ensure that dynamics reached the steady-state. We then computed the LFD from the second half of the simulated data.}

\RV{To compare the fits to the data of the SLRM and SLM, the same number (30) of logarithmically-spaced values of $\tau^{\prime}$ (SLM) and $\tau$ (SLRM) were scanned for each system, and the JSD between the empirical and model LFDs were computed. The particular range of $\tau^{\prime}$ and $\tau$ values scanned differed between the three systems; they were selected to ensure that a minima in JSD existed between the bounding values. The bounding values of this interval were $10, 10^4$ years for cities; $0.08, 100$ days for the microbiome; and $2.5, 5 \times 10^4$ years for forests. We computed the JSD on binned data when comparing the LFD of the SLRM and SLM with the observed data. The binning was determined using the Freedman-Diaconis estimator on the observed data. Note that since we do not have an analytical prediction for the SLM, we cannot directly compute the likelihood of the observed data to measure a. The JSD, on the other hand, can be computed between the binned observed and predicted distributions. }

\subsection*{Environmental stochasticity \RV{can produce} normally distributed fluctuations}
The dynamics of a population driven purely by environmental noise is given by 
\begin{equation}
\frac{dx_i}{dt} =  E_i x_i, \eta_i(t),
\end{equation}
where $E_i$ quantifies the strength of environmental noise and $\eta_i$ is \RV{delta correlated Gaussian noise or white noise}~\cite{lande_stochastic_2003,fung_reproducing_2016}. We can rewrite this equation for $\log x_i$ instead as 
\begin{equation}
\frac{d \log x_i}{dt} =  E_i \, \eta_i(t).
\end{equation}
Clearly, the fluctuations of $\log x_i$ are now \RV{normally} distributed. Thus environmental noise can produce a normally distributed LFD. Note, however, that this equation does not have a steady-state. Additional terms are required in the dynamical equation to stabilize the population and ensure a steady-state. 

Although a subset of microbes have an LFD that is well-explained by a normal distribution, we are able to fit the majority of the data without using environmental noise. While adding environmental noise would affect model behavior, the success of the model without environmental noise suggests that the populations could be in a parameter-range where the effect of environmental noise is unimportant or that the quantities we examine are not sensitive to the addition of environmental noise on top of demographic noise.

\RV{\subsection*{Data availability:} All datasets analyzed in this manuscript are publicly available. Code used to process and analyze the data as described is available on Github (\url{https://github.com/ashish-b-george/Universal-fluctuations})~\cite{b_george_universal-fluctuations_2023}. Employment data at the county level was obtained from the Quarterly Census of Employment and Wages from the U.S. Bureau of Labor Statistics (\url{https://www.bls.gov/cew/downloadable-data-files.htm}). Microbiome data from Ref.~\cite{caporaso_moving_2011} was obtained and processed as in Ref.~\cite{ji_macroecological_2020}. Forest data was obtained from the Center for Tropical Forest Science website~\cite{condit_complete_2019}.}

\subsection*{Acknowledgements}
The authors would like to thank Zachary Miller, Alice Doucet Beaupr\'{e}, and members of the O'Dwyer group for helpful feedback and comments. The authors acknowledge funding support from Simons Foundation Grant \#376199 and McDonnell Foundation Grant \#220020439 to J.O.D. The BCI forest dynamics research project was made possible by NSF grants to S.P. Hubbell: DEB \#0640386, DEB \#0425651, DEB \#0346488, DEB \#0129874, DEB \#00753102, DEB \#9909347, DEB \#9615226, DEB \#9405933, DEB \#9221033, DEB \#9100058, DEB \#8906869, DEB \#8605042, DEB \#8206992, DEB \#7922197, support from CTFS, the Smithsonian Tropical Research Institute, the John D. and Catherine T. MacArthur Foundation, the Mellon Foundation, the Small World Institute Fund, and numerous private individuals, and through the hard work of over 100 people from 10 countries over the past two decades. The plot project is part the Center for Tropical Forest Science, a global network of large-scale demographic tree plots.



\bibliography{bib_cities,refs}
\bibliographystyle{naturemag}
\end{document}


\title{Supplementary Information: Universal abundance fluctuations across microbial communities, tropical forests, and urban populations}
\author{Ashish B. George and James O'Dwyer }
\maketitle
\tableofcontents
\renewcommand{\theequation}{S\arabic{equation}}
\renewcommand{\thefigure}{S\arabic{figure}}
\renewcommand{\thesection}{S\arabic{section}}
\renewcommand{\thetable}{S\arabic{table}}
\setcounter{equation}{0}
\setcounter{figure}{0}
\setcounter{section}{0}
\setcounter{table}{0}

\section{Solving the Stochastic Linear-Response Model}
The stochastic Linear-Response model is given by
\begin{equation}
\frac{dx_i}{dt} =\frac{x^*_i}{\tau_i}  -  \frac{x_i}{\tau_i}     + \sqrt{ 2 \sigma_i x_i}\, \eta_i(t),
\label{eq:dxdt}
\end{equation}
where $x^*_i$ is the equilibrium abundance, $\tau_i$ sets timescale of the deterministic forces, $\sigma_i$ determines the strength of population fluctuations, and $\eta(t)$ is delta-correlated \RV{Gaussian noise or white noise} ($\langle \eta(t) \eta(t\prime) \rangle = \delta (t-t\prime)$).

The Fokker-Planck equation for the probability density function corresponding to the above stochastic differential equation, by the Ito prescription,  
\begin{equation}
\frac{\partial  P_i(x,t)}{\partial t} = -\frac{\partial }{\partial x} \left[  \left(\frac{x^*_i-x}{\tau_i}  \right)  P_i(x,t)  \right]  + \sigma_i \frac{\partial^2 }{\partial x^2} \left[ x P_i(x,t) \right] 
\end{equation}

\subsection{Steady-state solution}

At steady-state, the time derivative is zero, and so we get an equation for the steady-state probability density function $P_{\mathrm{ss},i}(x)$ :

\begin{equation}
\left(\frac{x^*_i-x}{\tau_i}  \right)  P_{\mathrm{ss},i} (x)=   \sigma_i \frac{\partial }{\partial x} \left[ x P_{\mathrm{ss},i} (x) \right]. 
\end{equation}

To solve this equation, we define $Q_i(x)=x P_{\mathrm{ss},i} (x)$ and simplify to get
\begin{equation}
\frac{1}{Q_i(x)} \frac{\partial Q_i(x) }{\partial x} = \frac{1}{\sigma_i \tau_i}  \left(\frac{x_i^*}{x} -1\right).
\end{equation}

This equation can be integrated to get
\RV{\begin{equation}
Q_i(x)= c x^{\frac{x_i^*}{\sigma_i \tau_i}} e^{-\frac{x}{\sigma_i \tau_i}},
\end{equation}}
where $c$ is an undetermined constant of integration. We can now calculate the steady-state pdf using $P_{\mathrm{ss},i} (x)= x^{-1}Q_i(x) $ and calculate the constant of integration from the normalization condition $\int_{-\infty}^{\infty} P_{\mathrm{ss},i} (x) dx =1$. This gives us the result

\begin{equation}
P_{\mathrm{ss}, i}(x) = \frac{(\sigma_i\tau_i)^{-x^*_i/\sigma_i \tau_i} }{\Gamma(x^*_i/\sigma_i \tau_i)} x^{\frac{x^*_i}{\sigma_i \tau_i}-1} e^{-\frac{x}{\sigma_i\tau_i}},
\label{eq:p_ss}
\end{equation}
which is the Gamma distribution. The properties of the Gamma distribution are well-known. The mean of the Gamma distribution is $x^*_i$ and the ratio of variance to mean-squared is the ENR $\frac{x^*_i}{\sigma_i \tau_i}$.

\subsection{Time-dependent solution}
We can also obtain the full time-dependent solution of the SLRM. Analytical solutions of this model have been obtained in the context of diffusion processes, bond interest rates, and birth-death models~\cite{feller_two_1951, cox_theory_1985, azaele_dynamical_2006}. To obtain the solution, we consider the backward Kolmogorov equation describing the dynamics

\begin{equation}
-\frac{\partial  P_i(x,t)}{\partial t} = \left(\frac{x^*_i-x}{\tau_i}  \right)  \frac{\partial }{\partial x}  P_i(x,t)  + \sigma_i x \frac{\partial^2 }{\partial x^2}  P_i(x,t) 
\end{equation}

We can non-dimensionalize the parameters by changing variable to $z=\frac{x}{\sigma_i \tau_i}$, $\theta= \frac{t}{\tau_i}$ to get
\begin{equation}
-\frac{\partial  P_i(z,\theta)}{\partial \theta} = \left(\frac{x^*_i}{\sigma_i \tau_i} - z \right)  \frac{\partial }{\partial z}  P_i(z,\theta)  +  z \frac{\partial^2 }{\partial z^2}  P_i(z,\theta) 
\end{equation}

We now use the separation of variables ansatz to look for solutions of the form $P(z,\theta)= q_{\lambda}(z) r_{\lambda}(\theta)$. Plugging this in, we get
\begin{equation}
\frac{-1}{r_{\lambda} (\theta)}  \frac{\partial  r_{\lambda}(\theta)}{\partial \theta} = \frac{1}{q_{\lambda}(z)} \left[  \left(\frac{x^*_i}{\sigma_i \tau_i} - z \right)  \frac{\partial q_{\lambda}(z) }{\partial z}    +  z \frac{\partial^2 q_{\lambda}(z)}{\partial z^2}   \right] =-\lambda.
\end{equation}
Hence we have $r_{\lambda}(\theta)= e^{\lambda \theta}$ and an equation for $q_{\lambda}(z)$:
\begin{equation}
\left(\frac{x^*_i}{\sigma_i \tau_i} - z \right)  \frac{\partial q_{\lambda}(z) }{\partial z}    +  z \frac{\partial^2 q_{\lambda}(z)}{\partial z^2}   + \lambda q_{\lambda}(z)=0.
\end{equation}

This is the confluent hypergeometric equation~\cite{abramowitz_handbook_2013}. It has two linearly independent solutions when $\frac{x^*_i}{\sigma_i \tau_i}$ is positive, denoted as $\psi_1(z)$ and $\psi_2(z)$. Hence the general solution is of the form $q_{\lambda}(z)= c_1 \psi_1(z) + c_2 \psi_2(z) $.

The two functions are defined as $\psi_1(z)= U(-\lambda,\frac{x^*_i}{\sigma_i \tau_i};z)$ and $\psi_2(z)= e^z U(\lambda+ \frac{x^*_i}{\sigma_i \tau_i},\frac{x^*_i}{\sigma_i \tau_i};z)$, where $U(a,b;z)$ is defined by the integral $U(a,b;z)= \frac{1}{\Gamma(a)} \int_0^{\infty} e^{-z s} s^{a-1}(1+a)^{b-a-1} ds$.

Imposing finite moments for the solution sets $c_2=0$ and $\lambda$ to take integer values. At these integer values, the eigen functions are Laguerre polynomials. Following Ref.~\cite{azaele_dynamical_2006}, we can solve for the coefficients assuming a reflecting boundary condition at $z=0$ and delta function initial condition at $t=0$, $\delta(x-x_0)$.

The time-dependent solution thus reads,
\begin{multline}
    P_i(x,t|x_0,0)= \frac{1}{\sigma_i \tau_i \left( 1- e^{-t/\tau_i}\right)}   \left( \frac{x}{x_0 e^{-t/\tau_i}}\right)^{-\frac{1}{2} + \frac{x_i^*}{2 \sigma_i \tau_i} }\\ \mathrm{exp}\left[- \frac{x+x_0 e^{-t/\tau_i} }{\sigma_i \tau_i \left( 1- e^{-t/\tau_i}\right)}\right]  I_{\frac{x_i^*}{ \sigma_i \tau_i}-1}  \left[ \frac{2 \sqrt{x x_0 e^{-t/\tau_i} }}{\sigma_i \tau_i \left( 1- e^{-t/\tau_i}\right) }   \right].
\label{eq:time_dep_sol}
\end{multline}
This is, in fact, the non-central chi-squared distribution, \RW{$\chi^2( \frac{2x}{  \sigma_i \tau_i \left( 1- e^{-t/\tau_i}\right) };\frac{2x_i^*}{ \sigma_i \tau_i}, \frac{2 x_0 e^{-t/\tau_i}}{  \sigma_i \tau_i \left( 1- e^{-t/\tau_i}\right) } )$}.


\subsection{The distributions of fold-change and logarithmic fold-change}

We can use the time-dependent solution ~(Eq.~\eqref{eq:time_dep_sol}) and the steady-state distribution~((Eq.~\eqref{eq:p_ss}) to calculate the probability of abundance fluctuations. Specifically, we will first calculate the probability distribution of the fold-change in a time interval $\Delta t$, $\rho= x(\Delta t)/x(0)$.

The fold-change distribution,  $P_{\mathrm{fc},i} (\rho, \Delta t)$, is given by
\begin{equation}
    P_{\mathrm{fluc},i} (\rho, \Delta t) = \int_{0}^{\infty} d x(0) \int_{0}^{\infty} dx(\Delta t) P_i(x(\Delta t),t+\Delta t| x(0),0 ) P_{ss}(x(0)) \delta\left(\rho-\frac{x(\Delta t)}{x(0)}  \right).
\end{equation}

Plugging the expressions from Eqs.~\eqref{eq:time_dep_sol} and ~\eqref{eq:p_ss} in the above, the integral is evaluated as in Ref.~\cite{azaele_dynamical_2006} to get 
\RW{
\begin{multline}
    P_{\mathrm{fc},i} (\rho, \Delta t) = \frac{2^{\frac{x_i^*}{\sigma_i \tau_i}-1} }{\sqrt{\pi}  }
    \frac{\Gamma\left( \frac{x_i^*}{\sigma_i \tau_i} +\frac{1}{2} \right) }{\Gamma\left( \frac{x_i^*}{\sigma_i \tau_i} \right)  }
    \frac{ (\rho+1) }{\rho}  
    \frac{ \left(e^{ \Delta t/\tau_i}\right)^{\frac{x_i^*}{2\sigma_i \tau_i}}}{1-e^{- \Delta t/\tau_i}}  \\
    \left( \frac{\sinh\left( \frac{\Delta t}{2 \tau_i} \right) }{\rho} \right)^{\frac{x_i^*}{\sigma_i \tau_i}+1}
    \left(  \frac{4 \rho^2}{ (\rho+1)^2 e^{\Delta t/\tau_i}-4\rho  }    \right)^{\frac{x_i^*}{\sigma_i \tau_i}+\frac{1}{2}}
\label{eq:fc_dist}
\end{multline}}

We can now change variables from the fold-change $\rho= \frac{x(\Delta t)}{x(0)}$ to logarithmic fold-change $\lambda=  \ln \frac{x(\Delta t)}{x(0)}$ by using $P_{\mathrm{fluc},i} (\lambda, \Delta t)= e^{\lambda}   P_{\mathrm{fc},i} (e^{\lambda}, \Delta t) $. Simplifying, we get the expression for the Logarithmic Fold-change distribution (LFD),

\begin{multline}
P_{\mathrm{fluc},i}(\lambda, \Delta t ) =  \frac{\Gamma\left(\frac{x^*_i}{\sigma_i \tau_i} +\frac{1}{2} \right)}{ 2\sqrt{\pi}  \Gamma\left(\frac{x^*_i}{\sigma_i \tau_i}  \right)} \left(e^{\frac{\Delta t}{\tau_i} } -1\right)^{\frac{x^*_i}{\sigma_i \tau_i}} e^{\frac{\Delta t}{2 \tau_i} }  \\
\cosh{\frac{\lambda}{2}}  \left( \frac{1 }{ e^{\frac{\Delta t}{\tau_i}} (\cosh{\frac{\lambda}{2}})^2  -1 }  \right)^{\frac{x^*_i}{\sigma_i \tau_i} +\frac{1}{2}}.
\label{eq:p_fluc}
\end{multline}

The corresponding cumulative distribution function can be computed as well, to make sampling the distribution easier. The cumulative distribution function corresponding to  $P_{\mathrm{fluc},i}(\lambda, \Delta t )$,  $CDF_{\mathrm{fluc},i}(\lambda, \Delta t )$ is given by

\begin{multline}
CDF_{\mathrm{fluc},i}(\lambda, \Delta t ) = \frac{1}{2}+  \frac{\Gamma\left(\frac{x^*_i}{\sigma_i \tau_i} +\frac{1}{2} \right)}{\sqrt{\pi} \Gamma\left(\frac{x^*_i}{\sigma_i \tau_i}  \right)} 
 \sinh{\frac{\lambda}{2}} \left(1-e^{-\frac{\Delta t}{\tau_i}} \right)^{ -\frac{1}{2}} 
 \prescript{}{2}{\mathrm{F}_1} \left[\frac{1}{2}; \frac{x^*_i}{\sigma_i \tau_i} + \frac{1}{2}; \frac{3}{2}; \frac{- \sinh^2{\frac{\lambda}{2}} }{1-e^{-\frac{\Delta t}{\tau_i}}} \right]
\end{multline}

where $\prescript{}{2}{\mathrm{F}_1} $ denotes the hypergeometric function.

\section{The SLRM as a linearization of a nonlinear model}
The most straightforward interpretation of the SLRM is that it arises from a more complicated nonlinear model as a linearization around the model equilibrium. Many ecological population models converge to a unique stable equilibrium~(see Table 1 in ~\cite{george_ecological_2023}). If we assume that the deterministic part of the dynamics is described by an equation $\frac{dx_i}{dt} =  f(\vec{x})$, where $\vec{x}$ is the vector of species abundances in the system and has an equilibrium at $\vec{x}=\vec{x}^*$ . Then deviations from the equilibrium gives rise to a linear restoring force of the form $ \nabla f(\vec{x}^*) \cdot \left( \vec{x}^*-\vec{x}\right)$, where $\nabla$ refers to the gradient operator. The SLRM is obtained when we assume that the restoring force  to equilibrium on species $i$ is dominated by its own deviation from equilibrium i.e., we neglect contributions from coupling between species. \RV{Neglecting interactions allows us to restrict an otherwise large number of model parameters.}  

This does suggest a natural extension of the SLRM that accounts for species interactions, as discussed in the main text. A simple way to incorporate species interactions would be to augment the deterministic part of the SLRM with contributions from the other species  $ \nabla f(\vec{x}^*) \cdot \left( \vec{x}^*-\vec{x}\right)$. \RW{This would result in an extension of the SLRM described by
\begin{equation}
\frac{dx_i}{dt} =\nabla f(\vec{x}^*) \cdot \left( \vec{x}^*-\vec{x}\right)    + \sqrt{ 2 \sigma_i x_i}\, \eta_i(t).
\end{equation}
Investigating this model and applying it to data is an interesting direction for future work. }

\subsection{\RV{Example of SLRM arising from a model with interactions}}
\RV{Here we show an example scenario of how the SLRM might arise from a population dynamics model with interactions without linearizing the dynamics. Specifically, we consider a generalized Lotka-Volterra (gLV) model with immigration and demographic noise: 
\begin{equation}
\frac{dx_i}{dt} = m_i + g_i x_i - \sum_j A_{ij} x_i x_j + \sqrt{ 2 \sigma_i x_i}\, \eta(t), 
\end{equation}
where $m_i$ is the immigration, $g_i$ is the growth rate, the matrix $A$ measures inter-species interactions, $\sigma_i$ measures strength of demographic noise, and $\eta(t)$ is delta-correlated \RV{Gaussian} noise.}

\RV{The limit we consider, reminiscent of a mean-field approximation in physics, is obtained by replacing the inter-species interaction $A_{ij}$ with an average interaction strength $ B_i=  \frac{1}{S} \sum_j A_{ij}$, where $S$ is the number of species. This leads to a low-rank approximation of the interaction matrix, with model dynamics being described by
\begin{equation}
\frac{dx_i}{dt} = m_i + g_i x_i - B_i x_i \sum_j x_j + \sqrt{ 2 \sigma_i x_i}\, \eta(t).
\end{equation}}

\RV{For relative abundances, $\sum_j x_j =1$ and neglect noise correlations in the high diversity limit (alternatively for absolute abundances, we neglect fluctuations in the total population size so that $\sum_j x_j=c$ is a constant). With this, we get 
\begin{equation}
\frac{dx_i}{dt} = m_i + g_i x_i - B_i x_i c + \sqrt{ 2 \sigma_i x_i}\, \eta(t), 
\end{equation}
which can easily be recast into form of the SLRM (Eq.~\eqref{eq:dxdt}) by re-expressing the parameters as $m_i=\frac{x^*_i}{\tau_i}$  and $ g_i- B_i c = -\frac{1}{\tau_i}$. Thus, we obtain the SLRM from the model with interactions in a special limiting scenario. Note that this is intended to serve only as an example scenario where SLRM is derived from an interacting model without linearization. Whether the SLRM may arise as an effective model of more complex interacting models in other scenarios is an interesting topic for future work.}


\section{The SLRM for relative vs absolute abundances.}

The SLRM equations in the main text were written in terms of the relative abundances $x_i$, which is also the data we analyze. Many common population models, however, are written in terms of the absolute abundance $n_i$ instead. Here we show that when fluctuations in the total population size can be neglected, the SLRM for relative abundances implies an SLRM for absolute abundances as well.

We start by assuming the absolute abundances of a sector/species, $i$, in a city/community $c$ is described by an SLRM. Specifically, the dynamics is given by 
\begin{equation}
\frac{dn_{ci}}{dt} =\frac{n^*_{ci}}{\tau_{ci}}  -  \frac{n_{ci}}{\tau_{ci}}     + \sqrt{ 2 \sigma^{(n)}_{ci} n_{ci}}\, \eta_{ci}(t),
\label{eq:dndt}
\end{equation}
where without loss of generality we have assumed that the demographic parameters depend on both $c$ and $i$. The equation for the relative abundances $x_{ci}=n_{ci}/N_c$, where $N_c$ is the total population of the city/community is then given by 
\begin{equation}
\frac{dx_{ci}}{dt} =\frac{x^*_{ci}}{\tau_{ci}}  -  \frac{x_{ci}}{\tau_{ci}}     + \sqrt{ 2 x_{ci} \sigma^{(n)}_{ci}/N_{c}}\, \eta_{ci}(t),
\end{equation}
if we neglect variation in the total population $\frac{dN_{c}}{dt}$. This variation in total abundances could be small due to some unmodelled global feedback that leads to the approximated cancellation of positive and negative fluctuations of many population categories. We can now redefine the noise term by using  $\sigma^{(x)}_{ci}= \sigma^{(n)}_{ci}/N_{c}$.  With this redefinition, the equation becomes
\RV{\begin{equation}
\frac{dx_{ci}}{dt} =\frac{x^*_{ci}}{\tau_{ci}}  -  \frac{x_{ci}}{\tau_{ci}}     + \sqrt{ 2 x_{ci} \sigma^{(x)}_{ci}}\, \eta_{ci}(t),
\label{eq:dfdt}
\end{equation}
which is identical to Eq.\eqref{eq:dndt}, except with $\sigma^{(x)}_{ci}$.} Thus an SLRM for relative abundances is equivalent to an SLRM for absolute abundances when the fluctuations in the total population size can be neglected. In general, these fluctuation may not be neglected and transforming the SLRM to absolute abundances would need to assume a dynamical equation for $\frac{dN_{c}}{dt}$. 

In the main text, we have made an additional simplifying assumption. \RV{When analyzing city data, we assumed that the parameters $\tau_{ci}$,$x^*_{ci}$, and $\sigma^{(x)}_{ci}$ depend only on the sector  $i$ and are independent of the city $c$. Analogous assumptions were applied to forest data. For microbiome analysis, we consider longitudinal data from only a single community $c$ and do not integrate data from different communities; hence such an assumption was not required. Also, note that assuming $\sigma^{(x)}_{ci}$,  depend only on the sector $i$ implies that $\sigma^{(n)}_{ci}/N_{c}$ is a constant independent of city size. This makes fluctuations important for big and small cities alike. }



\FloatBarrier

\section{\RV{Literature review}}
\RV{In this section, we summarize works examining the abundances and temporal fluctuations of the three complex populations we study.}

\subsection{\RV{Microbiome}}
\RV{In the context of the microbiome time-series data, Refs.~\cite{descheemaeker2020stochastic,grilli_macroecological_2020,zaoli_macroecological_2021} have proposed and applied the Stochastic Logistic Model (SLM) to microbiome data. The Stochastic Logistic Model is a three-parameter model that combines a nonlinear, deterministic part describing logistic growth with environmental stochasticity. \RW{In Fig.7 in the main text}, we compare the performance of the Stochastic Linear Response Model (SLRM) and the Stochastic Logistic Model (SLM) in explaining the statistical distributions of abundances and its fluctuations in all three systems. We found that the SLRM outperformed the SLM the majority of cases (17 out of the 18 employment sectors, 72 out of 85 microbial species, and 2 out of the 4 forest clusters).}

\RV{A number of studies have also fit microbiome data with more complex models such as Lotka-Volterra and consumer-resource models~\cite{marino2014mathematical,marsland_minimal_2020}. Ref.~\cite{ho_competition_2021} examined the Logarithmic Fold-change Distribution (LFD) in microbiome data, aggregated over all species. They fit the LFD with a Laplace distribution and used this to narrow the parameter space of consumer-resource models they were investigating. In Fig.~\ref{fig:LFD_AIC}, we compare the fits of the empirical LFD in all three systems Laplace and Gaussian candidate distributions. The SLRM outperformed the other distributions in majority of cases. Finally, Ref.~\cite{zapien2022effect} proposed a birth-death-migration model for microbiomes but do not apply it to data.}

\subsection{\RV{Forests}}

\JOD{Using tropical forest data to investigate population fluctuations was catalyzed by two key innovations: empirically, the development of whole-plot censuses over multiple years by the CTFS~\cite{anderson2015ctfs} provided data that can be used to analyze population fluctuations at scale. In the area of theoretical modeling, there was the development of the neutral theory of biodiversity~\cite{Hubbell1979,Hubbell2001,Volkov2003}, and its potential explanation of static patterns of diversity like the species abundance distribution and spatial turnover~\cite{Rosindell2008,Rosindell2011,odwyer2013eob,odwyer2018cross}. This naturally led to the question of whether the combination of competition for space and demographic noise could explain temporal patterns alongside spatial patterns. Azaele et al~\cite{azaele_dynamical_2006} used the SLRM formulation to test whether square-root (i.e. demographic-like) noise could potentially explain patterns of fluctuations in the Barro Colorado Island plot, treating all species as governed by the same parameter values. Our model follows a similar approach, though we allow for populations to take on different parameter values according to which of four height niches they belong to.} 

\JOD{Subsequent developments in the analysis of forest population fluctuations~\cite{Chisholm2014b,kalyuzhny2015neutral,fung2016reproducing} considered the possibility that environmental stochasticity~\cite{Leigh1981,engen1996population}, may provide a more comprehensive explanation. These models introduce independent, uncorrelated environmental noise for each species alongside demographic noise and competition for space, providing a better fit to several data sets than demographic noise alone, though at the conceptual expense of introducing unidentified environmental variables. We also note that the metrics used to assess the fit to fluctuation sizes differ from those used here and by Azaele et al, and here in our analysis, and in our own analysis we show that square-root noise alone may provide an adequate explanation of tropical forest fluctuations.}

\subsection{\RV{Cities}}
\RV{For the dynamics of sectoral employment in cities, work has been mostly limited to regional effects, sector-specific analyses, or understanding the effect of total city size on employment composition by sector in the city~\cite{clark_employment_1998, armington_job_2004,hong_universal_2020, bettencourt_professional_2015}. In a broader context, simple mathematical models have been successfully applied to explain the distribution of population sizes across cities and population migration patterns~\cite{gabaix_zipfs_1999,verbavatz_growth_2020,simini_universal_2012}.}

\section{\RV{Alternative goodness of fit tests}}
\RV{In addition to the likelihood based goodness fit test presented in Fig.4, we applied two alternative goodness of fit tests to the data: a test based on the Jensen-Shannon Distance (JSD)~\cite{endres_new_2003,virtanen_scipy_2020} and the Kolomogorov-Smirnov (KS) test
~\cite{virtanen_scipy_2020,massey_kolmogorov-smirnov_1951}. \\}

\RV{The Jensen-Shannon Distance (JSD) measures the distances between two probability distributions. The JSD-based test was performed similarly to the likelihood based test. 1000 Random samples of the same size of the fitted distribution were generated at the estimated parameters. Then the JSD between the observed data and theoretical distribution was compared to JSDs of the random samples and the theoretical distribution. We passed the fit if the JSD between observed data and theory was was smaller than at least 1\% of the simulated data (i.e., rejected at a 1\% threshold). 13 of 18 sectors, 43 of 85 microbial species, and 3 of 4 forest clusters passed the JSD test. Visualizing the abundance distributions of species that were rejected by the test, we found that rejection was sometimes driven by just one or two outlier data points where the species was recorded as being highly abundant (see Fig.~\ref{fig:SSD_gof}C,G). Whether this signifies the ground truth or measurement error is an open question.}

\RV{The one-sample KS test is a non-parametric test that compares the observed data with a reference distribution (theoretical prediction). It generates a p-value associated with each comparison. 15 of 18 sectors, 55 of 85 microbial species, and 3 of 4 forest clusters passed the KS test at a 1\% threshold.}

\section{\RV{Robustness to changing time interval of sampling}}

\RV{To test the robustness of our results, we investigated the effect of changing the time interval between consecutive data points over which the Log Fold-change was calculated, $\Delta t$. This could be performed for the microbiome and cities because there were enough time points in both these data sets to allow us to vary $\Delta t$. To implement the test, we first obtained the empirical LFD for different values of sampling interval $\Delta t$. The theoretical LFD, $P_{\mathrm{fluc}}$,  depends on two parameters, the ENR and $\Delta t/ \tau$. We obtained the ENR from fitting the empirical abundance distribution (SSD) and then fit the empirical LFD to obtain $\Delta t/ \tau$ as $\Delta t$ was varied.}

\RV{Fig.~\ref{fig:city_robust_Dt} shows the results of this analysis for employment data in cities. We found that the inferred parameter combination $\Delta t/ \tau$ increased approximately linearly with increasing $\Delta t$ as we expect (Fig.~\ref{fig:city_robust_Dt}A). In other words, the inferred parameter $\tau$ is approximately independent of the sampling interval $\Delta t$ indicating a robust fit (Fig.~\ref{fig:city_robust_Dt}B).}

\RV{In contrast to the employment data, analysis of the microbiome data showed that the inferred parameter combination $\Delta t/ \tau$ increased slower than linearly with $\Delta t$ (Fig.~\ref{fig:microbe_robust_Dt}A). Furthermore, at large $\Delta t$ the inferred parameters of a few species oscillated. This spurious behavior can be understood by examining how the predicted LFD $P_{\mathrm{fluc}}$ changes as we vary $\Delta t / \tau$ (Fig.~\ref{fig:microbe_robust_Dt}B). The distribution does not change much once $\Delta t$ approaches $\tau$ or $\frac{\Delta t}{\tau} \gtrapprox 1$. Intuitively, this happens because the system equilibrates over a time scale of $\tau$. The fold-change distribution thus remains approximately unchanged for timescales greater than $\tau$. Since the inferred equilibration timescale for microbes is not significantly larger than the sampling interval, this makes the parameter estimation difficult as we increase $\Delta t$.
Fig.~\ref{fig:microbe_robust_Dt}C-F shows the data corresponding to one of the microbes that exhibited a spurious oscillation in its inferred parameter. Small changes in the data led to large changes in the inferred parameter because $\frac{\Delta t}{\tau} \gtrapprox 1$.}

\section{\RV{Are forest clusters significantly different?}}
\RV{In our analysis of BCI data, we classified species into four distinct clusters based on their maximum height. This was done under the assumption that species with similar height have similar access to light, in terms of level, variability, or horizontal uniformity, and hence are likely to be described the similar parameters~\cite{terborgh1985vertical}.  We utilized the results from the quantitative analysis of BCI data in Ref.~\cite{dandrea_counting_2020} to assign species to four distinct height clusters: shrubs, understory treelets, midstory trees, and canopy trees. }

\RV{In this section, we examine how different the extent of dissimilarity in the inferred parameters and emergent behavior of the four different height clusters. For this, we conducted goodness of fit tests to see if the parameters of individual tree clusters are  differed significantly from each other. Specifically, for each pair of clusters, eg. ($c_1$=shrubs, $c_2$=canopy), we took the fit parameters of cluster $c_1$ and computed the likelihood and JSD metrics using the data from cluster $c_2$. We compared this to the likelihood and JSD metrics obtained from 100 random samples from the LFD and 1000 samples from the SSD. Note that ($c_1$=shrubs, $c_2$=canopy) is different from ($c_1$=canopy, $c_2$=shrubs) and so there are 12 such pairs we can test.}

\RV{At a 5\% threshold for the LFD, the likelihood and JSD tests rejected 3 and 6 pairs respectively. All of the rejected pairs involved comparing a shrub with one of the other three clusters. Fig.~\ref{fig:comparing_forest_clusters} A indicates that the primary difference us that shrubs have a significantly shorter equilibration timescale than the other clusters. This manifest in the observed Log Fold-change data as shrubs experiencing much larger fluctuations than the other three clusters (Fig.~\ref{fig:comparing_forest_clusters} B). This might be reflective of the shorter generation times of shrubs. From the fitted distributions shown, we also observe that the fitted parameters of midstory trees and canopy trees are highly similar.}

\RV{The distinction between shrubs and the other three clusters appeared in the test results of the SSD as well. At a 5\% threshold for the SSD, the likelihood and JSD tests rejected 3 and 6 pairs respectively. Shrubs were involved in 6 of the 9 rejected pairs. Thus, in terms of the emergent properties analyzed in this study, shrubs form the most distinct height cluster among the BCI forest species.}

\section{\RV{Fitting absolute abundance data}}
\RV{In our analyses so far, we used relative abundance data in all three systems. This was done for a few reasons. First, sequencing does not provide absolute abundance data in the microbiome data, and so this allows us to treat data from all three systems identically. Second, our definition of relative abundances in employment, as the employment in a sector in a city divided by the total employment in the city, allows us to ignore the large variation in the absolute sizes of cities ~\cite{gabaix_zipfs_1999}. Third, it allows us to ignore the effects of changing overall population sizes (Fig.~\ref{fig:pop_size_meanCV}A,B).}

\RV{Despite the trends in overall population sizes (Fig.~\ref{fig:pop_size_meanCV}A,B), the fluctuations in the logarithmic fold-change of species/sectors calculated as $\frac{n_i(t+\Delta t)} {n_i(t)}$ are still large compared to the mean (Fig.~\ref{fig:pop_size_meanCV}C,D). Therefore, we tried fitting the empirical LFD calculated from absolute abundance data of species in forests and sectoral employment in cities (Fig.~\ref{fig:LFD_absabu}). The SLRM was still able to fit the data well. It outperformed other candidate distributions in all employment sectors in the city data and 2 of the 4 forest clusters, as compared to outperforming candidate distributions in all employment sectors and all 4 forest clusters when using LFD from relative abundance data. This slight decrease in performance against candidate distributions for the forest data may indeed be due to the temporal drift of overall population sizes.}

\RV{We also tried to fit the abundance distribution to the SSD predicted by the SLRM for forest data. The SLRM was able to fit the data reasonably well~(Fig.~\ref{fig:SSD_absabu}). We did not fit the abundance distribution of absolute sectoral employment across cities because we expect the variation in population sizes across cities to play a dominant role in this analysis. City sizes are known to vary widely and display a power-law like behavior~\cite{gabaix_zipfs_1999}. Indeed, this was one of our motivations behind normalizing the sectoral employment in a city by the total employment in the same city.}

\section{Supplementary figures}
\FloatBarrier

\begin{figure*}
\includegraphics[width=\textwidth]{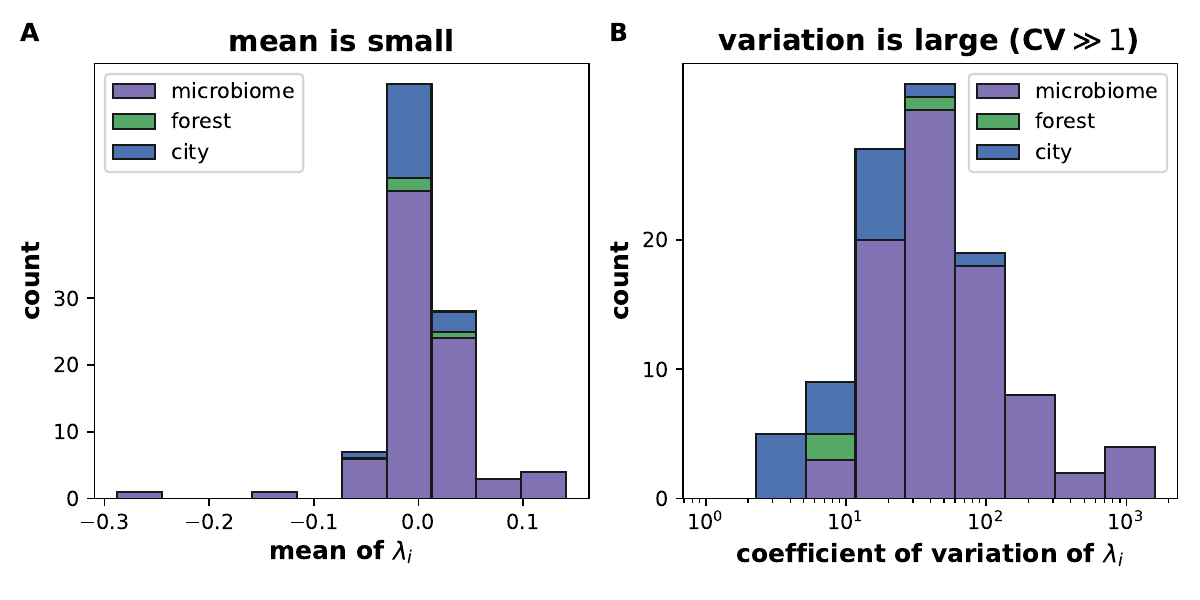}
\caption{\textbf{The mean logarithmic fold-change is much smaller than its standard deviation.} A) The mean logarithmic fold-change ($\lambda_i$) of each trajectory is smaller than the values of $\lambda_i$ shown in Fig.2. B) The coefficient of variation (ratio of standard deviation to mean) is larger than one in all data sets. This prompts us to neglect the small change in the mean in comparison to the large fluctuations between consecutive time points. Both panels show stacked histograms.}
\end{figure*}

\begin{figure*}
\includegraphics[width=\textwidth]{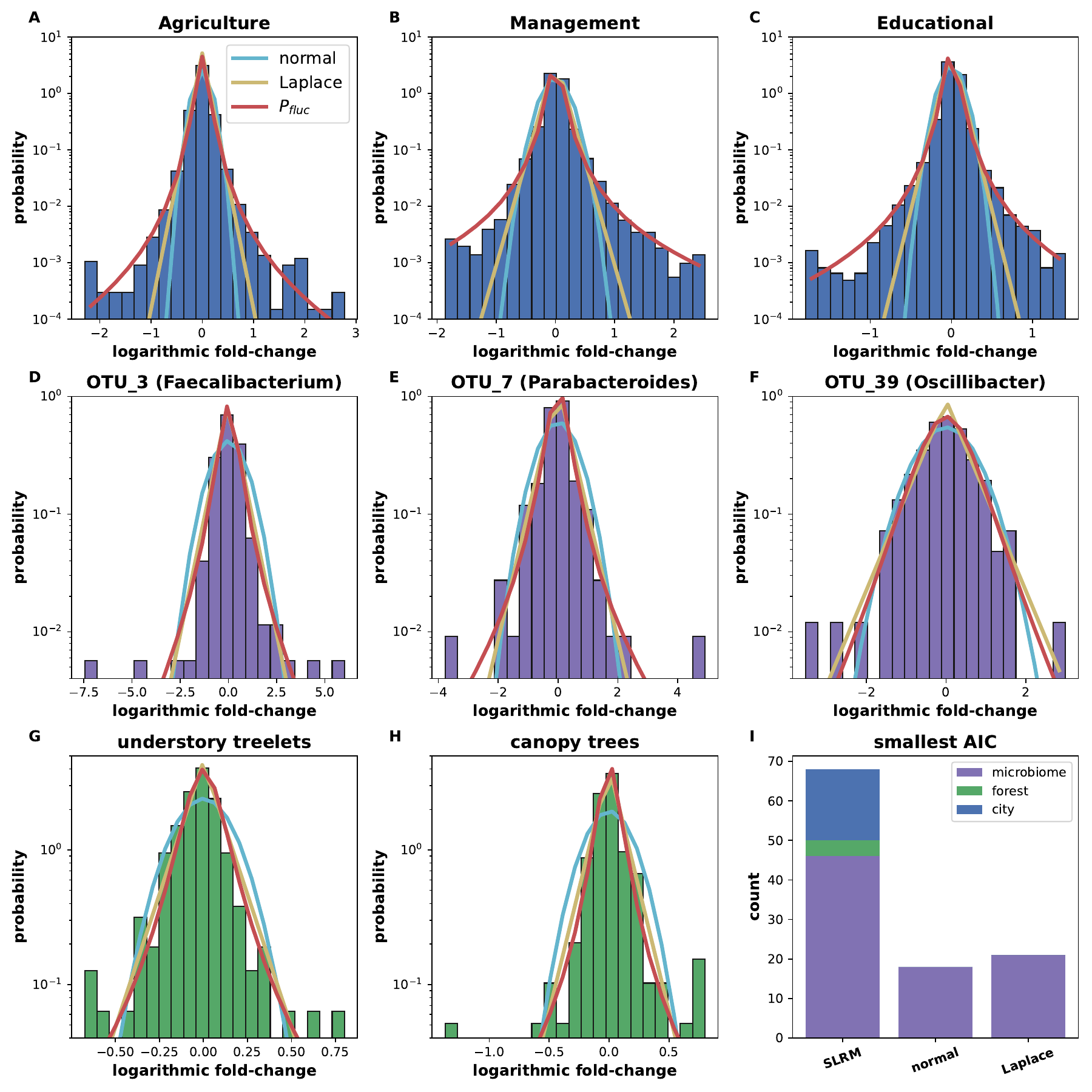}
\caption{\textbf{Model outperforms other candidate distributions in majority of cases.} The empirical Logarithmic Fold-change Distribution (LFD) is shown for three employment sectors \textbf{(A-C)}, three microbial species \textbf{(D-F)}, and two tree height clusters (\textbf{(G,H)}. The LFD is \RV{fit} by three candidate distributions: normal(cyan), Laplace (yellow), and the prediction from the SLRM, $P_{\mathrm{fluc}}$ (red). \textbf{(I)}The SLRM prediction has the smallest Akaike Information Criterion~\cite{parzen_information_1998} value in the majority of the cases, as show in the stacked histogram. Note that all fitting distributions assumed zero mean, in agreement with our observation of negligible drift (Fig.S1).}
\label{fig:LFD_AIC}
\end{figure*}

\begin{figure*}
\includegraphics[width=.5\textwidth]{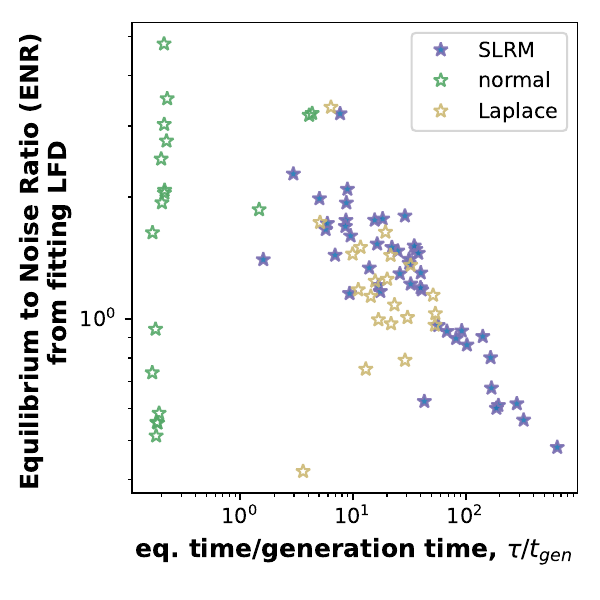}
\caption{\textbf{The microbial species labelled by the model best explaining their logarithmic fold-change distribution.} The SLRM fits the empirical LFD best in the majority of cases (46/85 fits), as measured by the Aikake Information Criterion.  Notably, all of the outliers at small $\tau$ values in Fig.3 are better fit by a normal distribution.}
\end{figure*}

\begin{figure*}
\includegraphics[width=.85\textwidth]{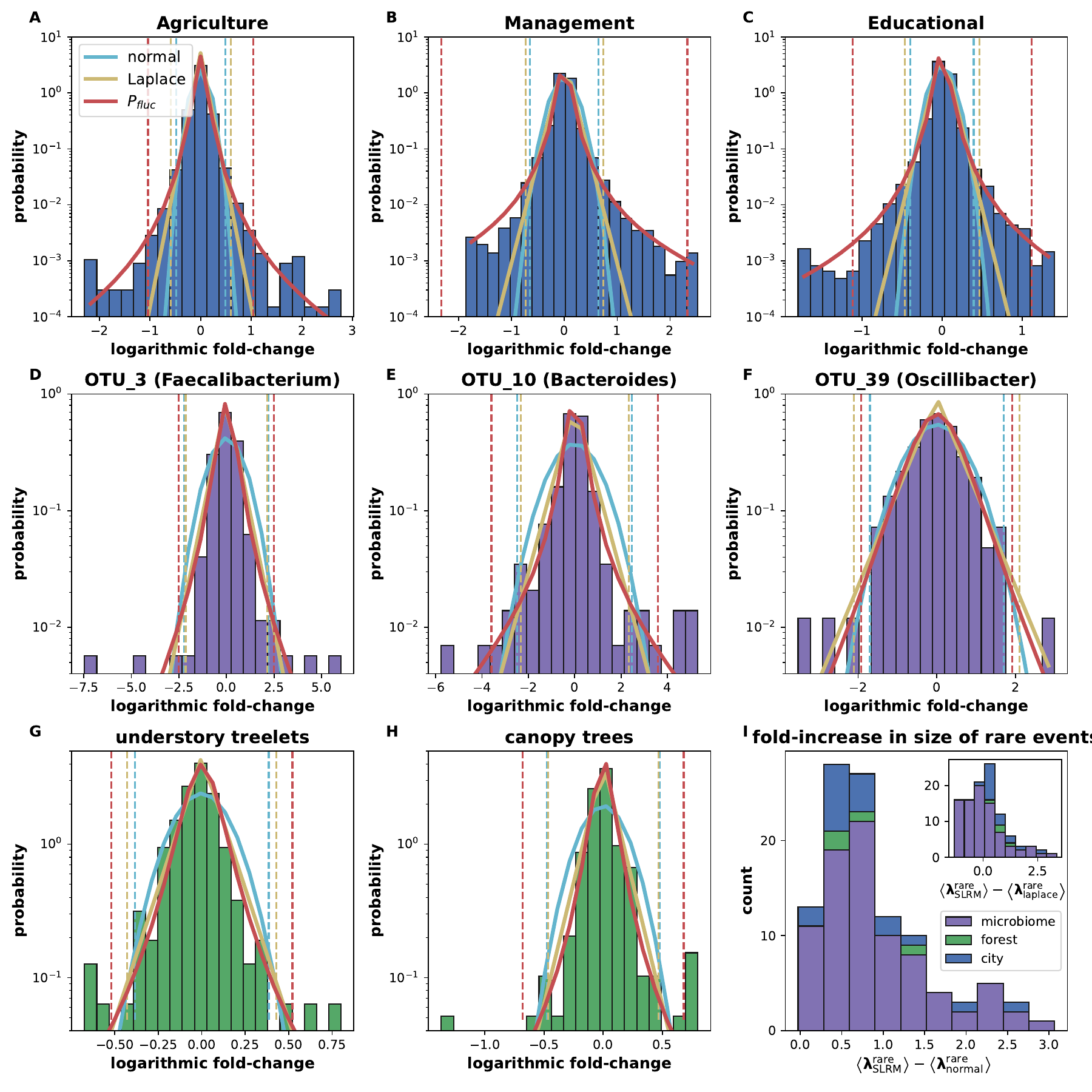}
\caption{\textbf{Difference in size of rare of events estimated by SLRM, Laplace and Normal distribution fits.} \textbf{A-H} Examples of Fits of the empirical LFD by the three candidate distributions: SLRM (red), Laplace (cyan), Normal (yellow). Dashed lines bound the central region where majority of the events take place (98\% for microbes and forests, 99.8\% for cities), as predicted by the three distributions (colored to match). The region outside the dashed lines illustrates the rare, large fluctuations. \textbf{I} The error in the estimated logarithmic fold-change of rare, large fluctuations when using alternative distributions instead of SLRM. The size of rare, large fluctuations estimated by each distribution is defined as the expected value of logarithmic fold-change $\lambda$ in the right tail of the distribution (to the right of the dashed lines in panels A-H), $\left \langle \lambda^{\mathrm{rare}}_{\mathrm{dist.}} \right \rangle$. The histogram shows the difference between the estimated size of large fluctuations by the SLRM $\left \langle \lambda^{\mathrm{rare}}_{\mathrm{SLRM}} \right \rangle$  and normal distribution $\left \langle \lambda^{\mathrm{rare}}_{\mathrm{normal}} \right \rangle$. The histogram demonstrates that we will consistently underestimate the size of rare events if we use a normal distribution for risk estimation. The inset shows the difference between the estimated size of large fluctuations by the SLRM $\left \langle \lambda^{\mathrm{rare}}_{\mathrm{SLRM}} \right \rangle$  and laplace distribution $\left \langle \lambda^{\mathrm{rare}}_{\mathrm{normal}} \right \rangle$. Rare events on the right side of the dashed line had 1\% chance of occurring for microbes and forests, and 0.1\% for cities. The expectation value of a rare event was defined as  $\left \langle \lambda^{\mathrm{rare}} \right \rangle = \int_{l}^{\infty} \lambda f(\lambda) d\lambda / \int_{l}^{\infty}  f(\lambda) d\lambda$ for probability distribution function $f(l)$ where the lower limit $l$ was defined as the $\lambda$ value demarcating a rare event (dashed line in panels A-H). }
\end{figure*}

\begin{table*}
\includegraphics[width=1\textwidth]{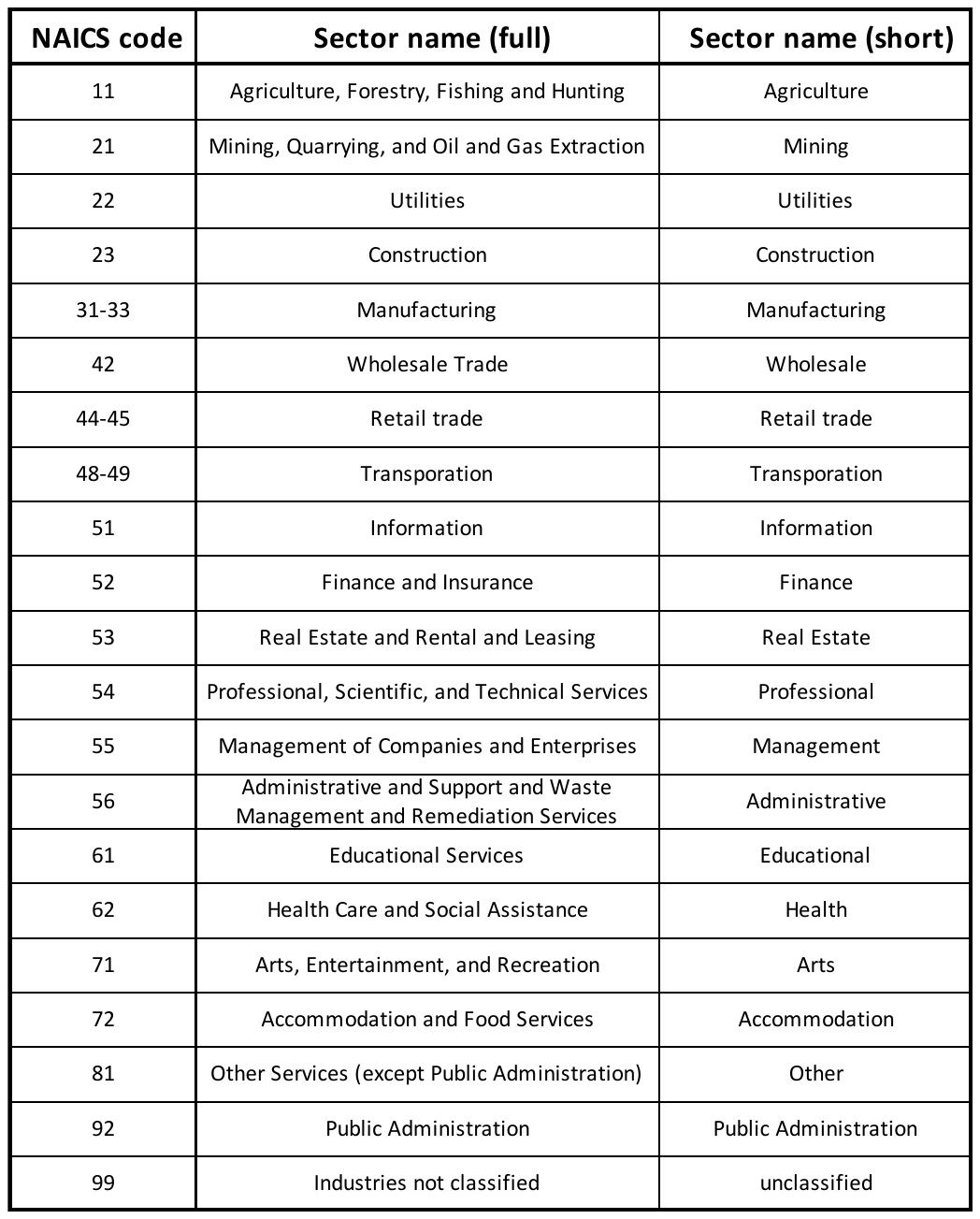}
\caption{\textbf{Table of NAICS categories at two digit resolution.} The shortened version of the NAICS category names are used in the text for clarity and brevity. }
\end{table*}

\begin{figure*}
\includegraphics[width=\textwidth]{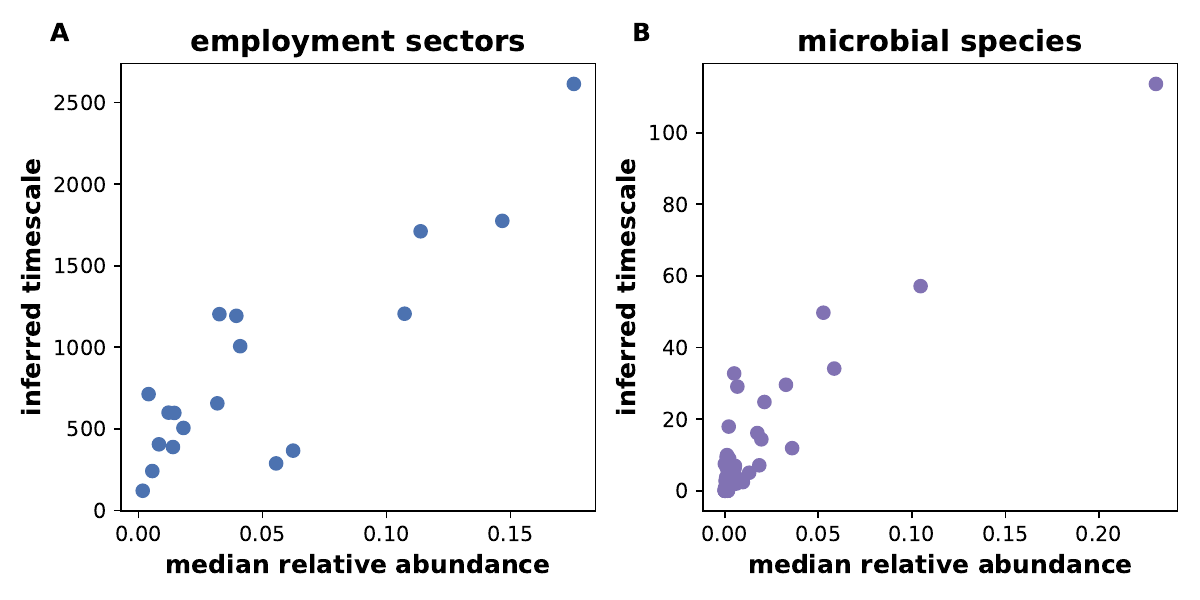}
\caption{\RV{\textbf{Inferred timescale, $\tau$ is correlated with abundance of the sector/species.} We found that the inferred timescale from fitting the empirical Log Fold-change Distribution (LFD) is correlated with median abundance of the employment sector \textbf{(A)} and microbial species \textbf{(B)}. Note that the LFD does not directly contain information about the relative abundance of a species/sector. The median relative employment is calculated across all cities at a given time-point (July 2016) and median relative species abundance is calculated across all time points for the microbiome. Calculated correlations were: pearson's r =$0.86$, spearman r =$0.66$ for cities and  pearson's r =$0.93$, spearman r =$0.68$ for the microbiome. All correlations were statistically significant $p<0.05$.  Axes are plotted on a linear scale because correlations were calculated on a linear scale.}}
\label{fig:tau_abu_corr}
\end{figure*}

\begin{figure*}
\includegraphics[width=\textwidth]{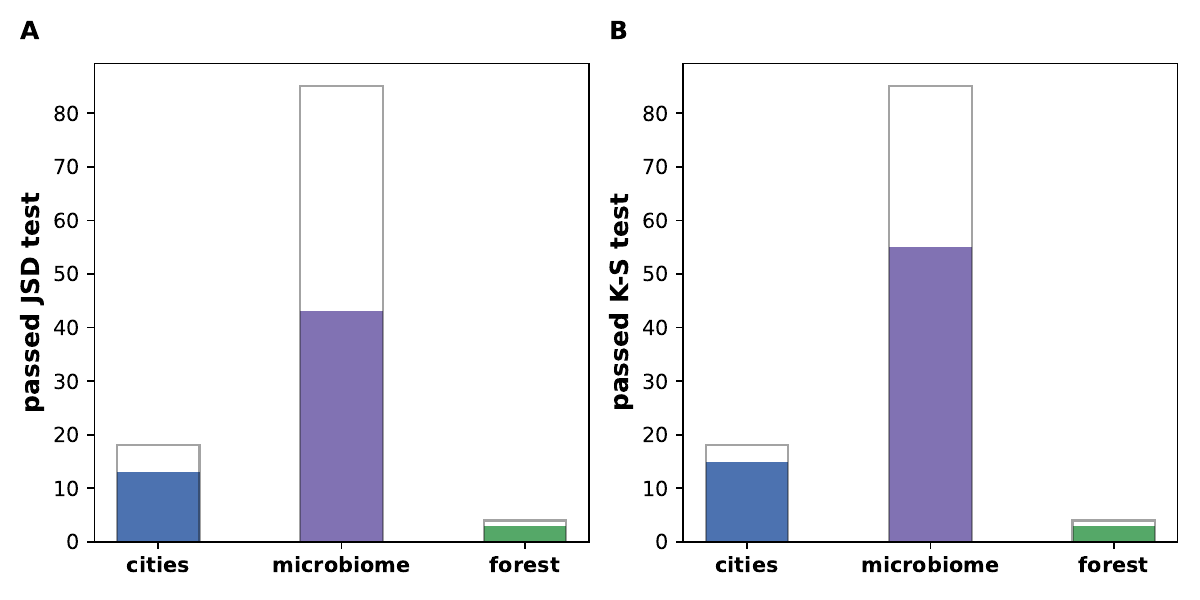}
\caption{\RV{\textbf{Alternative goodness of fit tests of the abundance distribution.} In addition to the likelihood based goodness fit test presented in Fig.4, we applied two alternative goodness of fit tests to the data: a test based on the Jensen-Shannon Distance (JSD)~\cite{endres_new_2003,virtanen_scipy_2020} and the Kolomogorov-Smirnov (KS) test~\cite{virtanen_scipy_2020,massey_kolmogorov-smirnov_1951}. A majority of species in each system passed the tests in each system. The species/sectors/clusters that passed the JSD test \textbf{(A)} and KS test \textbf{(B)} in each system are shown in solid colors. The background shows the total number of species in the systems. In particular, 15 of 18 sectors, 55 of 85 microbial species, and 3 of 4 forest clusters passed the KS test; 13 of 18 sectors, 43 of 85 microbial species, and 3 of 4 forest clusters passed the JSD test. }}
\label{fig:alt_gof_tests}
\end{figure*}

\begin{figure*}
\includegraphics[width=\textwidth]{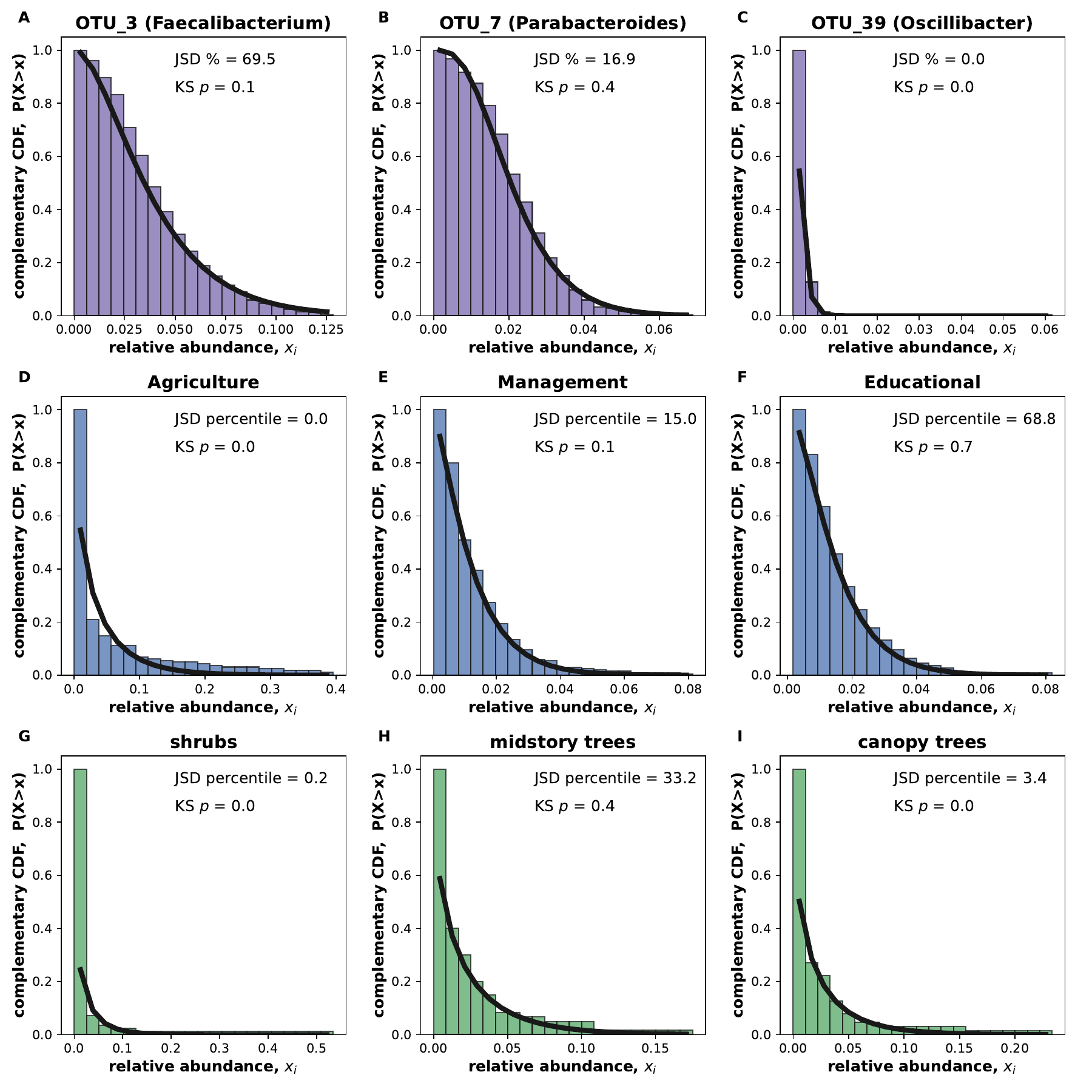}
\caption{\RV{\textbf{Abundance distributions with  goodness of fit test values.} \textbf{A-I)}The abundance distributions of a few species/sectors/clusters in each system. The percentile score in the JSD-based test and the p-value of the KS test are shown. The fit passed if the percentile score was $>1\%$ or if $p>0.01$. Panels C and G illustrate scenarios where the fit was rejected by a single outlier data point near 0.06 and 0.5 respectively.}}
\label{fig:SSD_gof}
\end{figure*}

\begin{figure*}
\includegraphics[width=\textwidth]{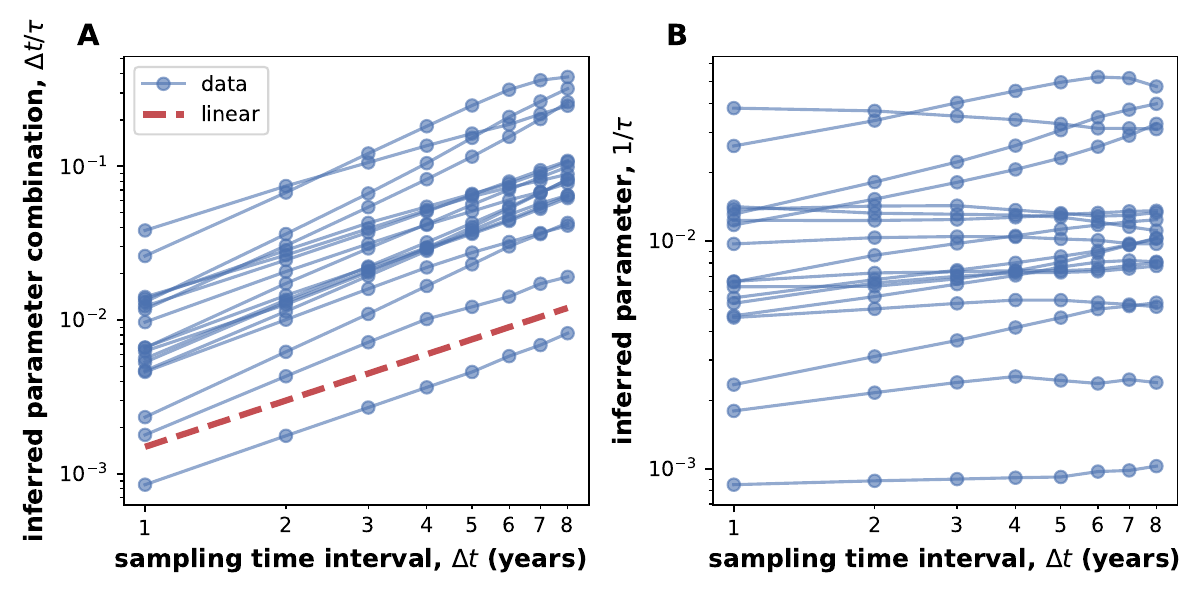}
\caption{\RV{\textbf{Scaling of inferred parameters with sampling interval for employment data.} We obtained the empirical LFD at different sampling intervals ($\Delta t$) and fit each one to the theoretical prediction $P_{\mathrm{fluc}}$.  \textbf{A)} The inferred parameter combination $\Delta t/ \tau$ from the fit increased linearly with the sampling interval as expected. \textbf{B)}In other words, the inferred timescale, $\tau$, remained approximately constant across fits using data at different $\Delta t$. Sampling intervals $\Delta t$ from 1 to 8 years were used to obtain the empirical LFD. Fits of the LFD were performed at a fixed ENR, which was obtained from fitting the abundance distribution.}}
\label{fig:city_robust_Dt}
\end{figure*}

\begin{figure*}
\includegraphics[width=.7\textwidth]{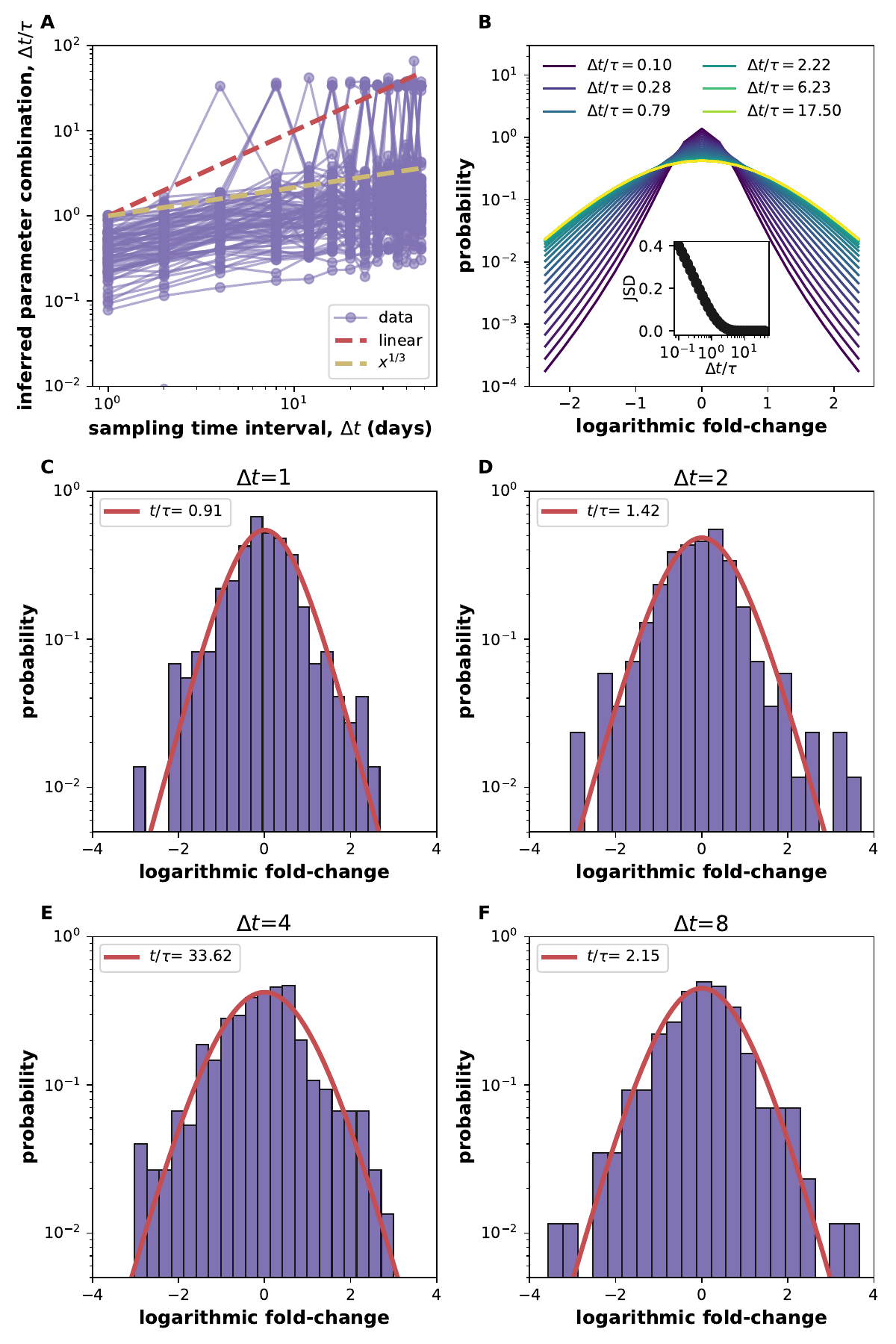}
\vskip -15pt
\caption{\RV{\textbf{Scaling of inferred parameters with sampling interval for microbiome data.} We obtained the empirical LFD at different sampling intervals ($\Delta t$) and fit each one to the theoretical prediction $P_{\mathrm{fluc}}$.  \textbf{A)} The inferred parameter combination $\Delta t/ \tau$ from the fit increased with the sampling interval, but slower than the expected (linear). The inferred parameters of a few species oscillated strangely, particularly at large $\Delta t$. 
\textbf{B)} The predicted distribution $P_{\mathrm{fluc}}$ changes very little for $\frac{\Delta t}{\tau} \gtrapprox 1$. This makes inferring $\Delta t/ \tau$ difficult once $\Delta t$ approaches $\tau$, which is the case for the microbiome. The inset shows the Jensen-Shannon Distances (JSD) between the distributions at the indicated $\Delta t/ \tau$ values and $\Delta t/ \tau =40$.The distance between these distributions vanishes quickly once $\Delta t/ \tau$ crosses 1.
\textbf{C-F)} The stiffness of the parameter inference procedure once $\Delta t \approx \tau$ causes the inferred parameter of a species to oscillate due to small changes in the underlying data. The panel shows the data for OTU 56, which had an ENR of $2.47$. The ENR in panel B was $2.5$. Fits of the LFD were performed at a fixed ENR, which was obtained from fitting the abundance distribution.}}
\label{fig:microbe_robust_Dt}
\end{figure*}

\begin{figure*}
\includegraphics[width=\textwidth]{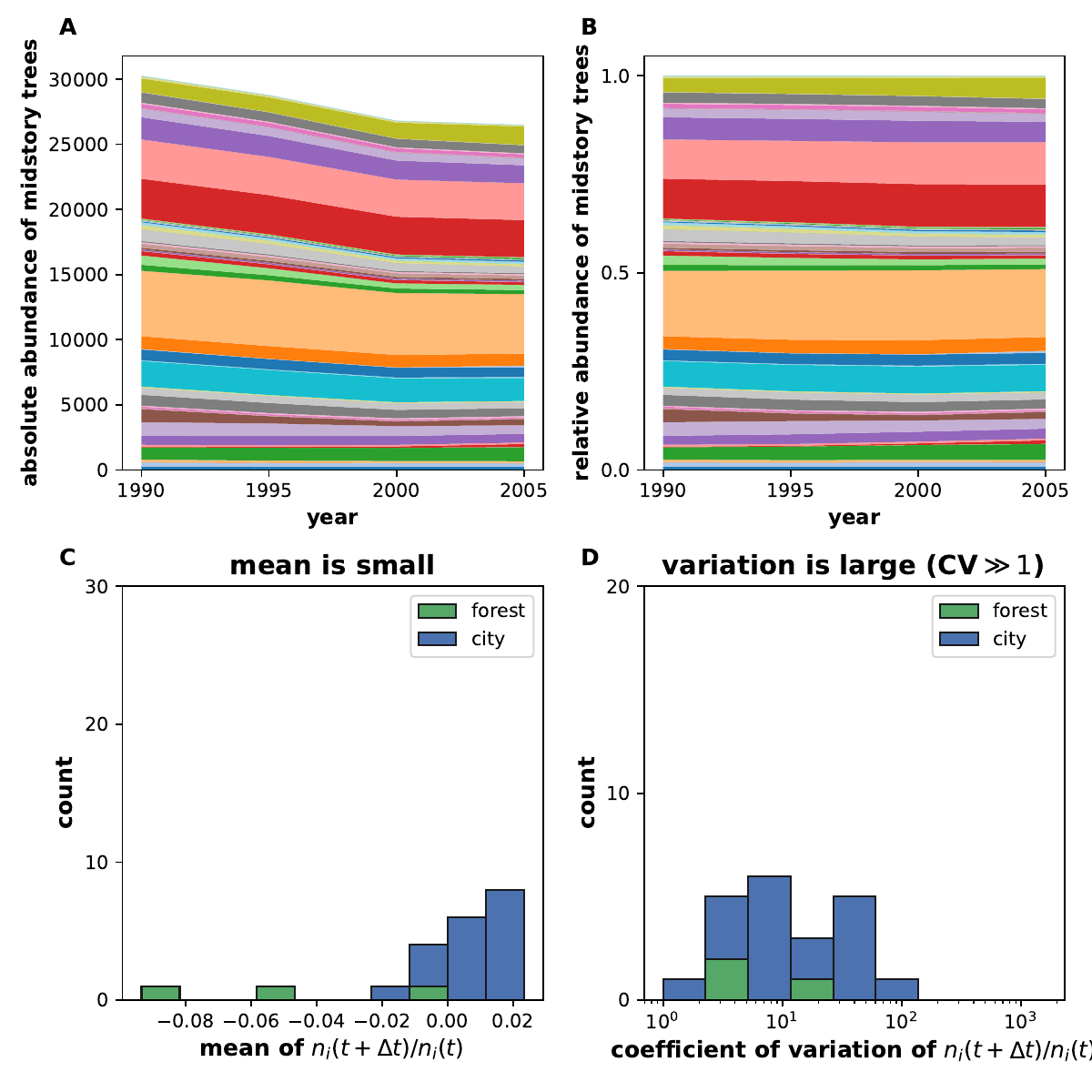}
\caption{\RV{\textbf{The temporal trends in absolute population sizes and the importance of fluctuations in logarithmic fold-change of absolute abundance data.} Absolute abundance data may display apparent trends over observation time scales. A) The total population sizes of tree clusters, such as midstory trees (shown), decreased over the observation time window. B) Studying the relative abundances allows us to disentangle the effects of changes in overall population sizes from the fluctuations of species within the overall population cluster. Colors denote different species within the cluster. C) The mean logarithmic fold-change computed from absolute abundance data, $\frac{n_i(t+\Delta t)} {n_i(t)}$, of each trajectory is smaller than the logarithmic fold-change values in Fig.~\ref{fig:LFD_absabu}. D) The coefficient of variation (ratio of standard deviation to mean) is larger than one in both data sets. This allows us to neglect the small change in the mean in comparison to the large fluctuations between consecutive time points, and fit our model to the empirical LFD in Fig.~\ref{fig:LFD_absabu}. Both panels show stacked histograms.}}
\label{fig:pop_size_meanCV}
\end{figure*}

\begin{figure*}
\includegraphics[width=\textwidth]{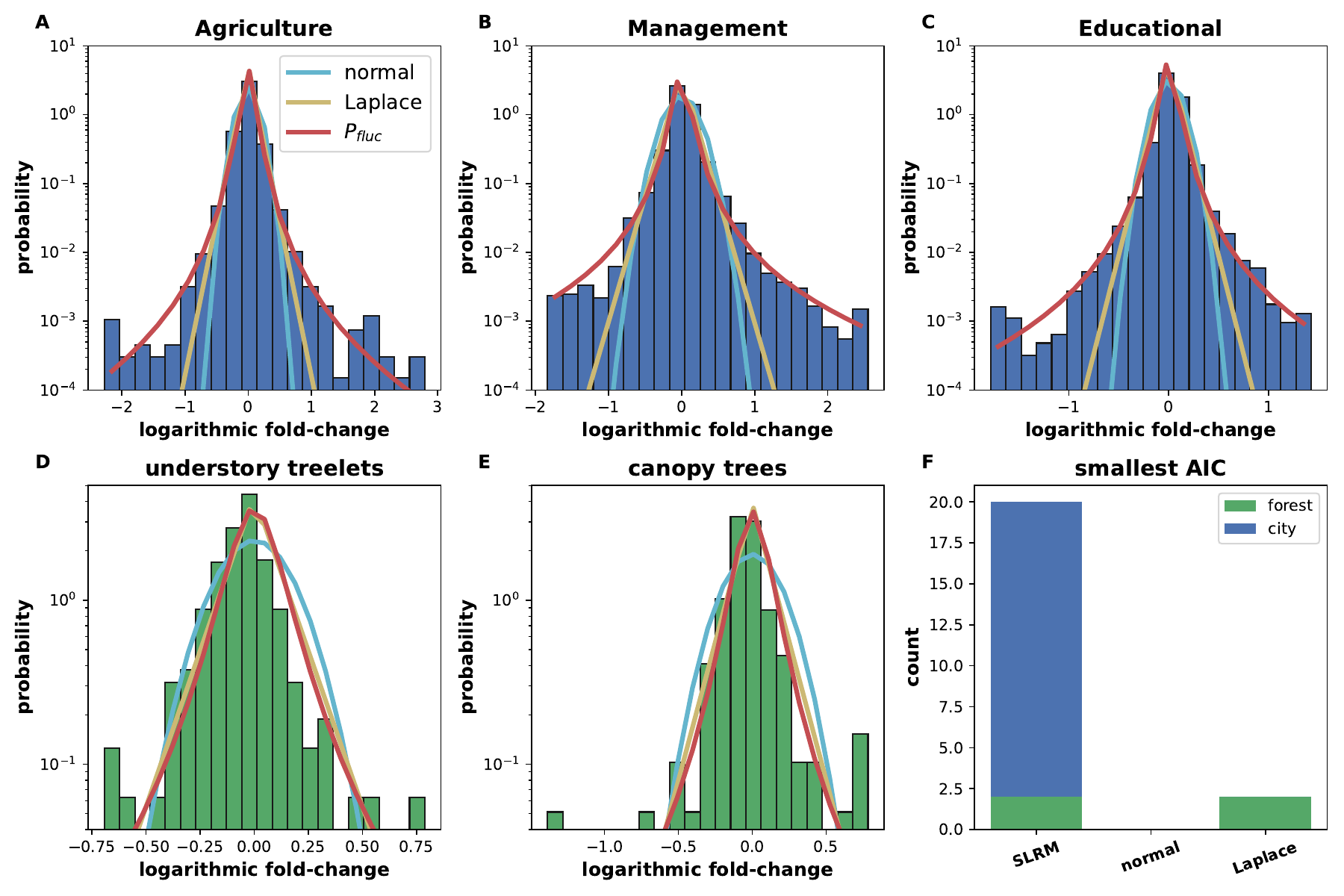}
\caption{\RV{\textbf{Fitting the LFD obtained from count data.} The empirical Logarithmic Fold-change Distribution (LFD) is computed using absolute abundances instead of relative abundances for cities and forests, where such data is available. The empirical LFD is shown for three employment sectors \textbf{(A-C)} and two tree height clusters (\textbf{(G,H)}. The LFD is fit by three candidate distributions: normal(cyan), Laplace (yellow), and the prediction from the SLRM, $P_{\mathrm{fluc}}$ (red). \textbf{(I)}The SLRM prediction has the smallest Akaike Information Criterion~\cite{parzen_information_1998} value in all of the employment data and 2 of the 4 forest clusters. When fitting LFD from absolute abundance data, temporal fluctuations in overall population size can contribute if they are large enough. This may explain why SLRM performed better than the other candidate distributions in all four forest clusters for relative abundances but only two clusters for absolute abundances.}}
\label{fig:LFD_absabu}
\end{figure*}

\begin{figure*}
\includegraphics[width=\textwidth]{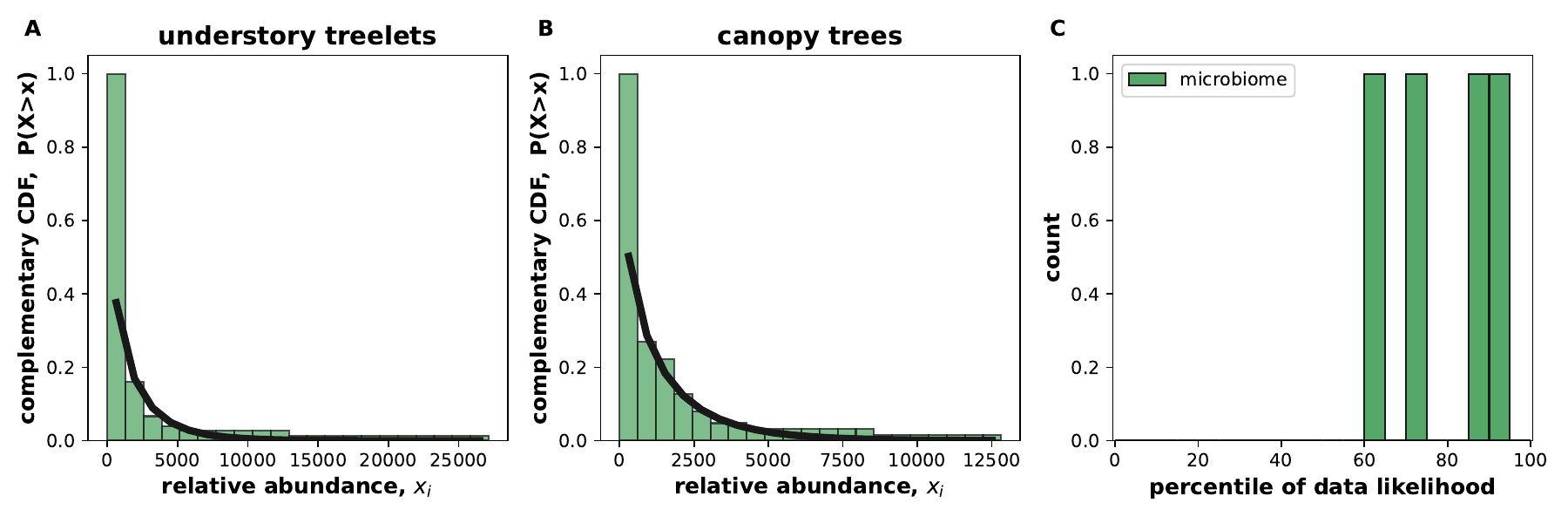}
\caption{\RV{\textbf{Fitting the SSD obtained from count data.} The empirical SSD (LFD) is computed using absolute abundances for forests. \textbf{A,B)}The empirical distribution of abundances of two tree height clusters and the fit by the model. \textbf{C)} All of the fits pass the likelihood-based goodness of fit test. The same analysis cannot be done for cities since we are considering different cities, with  different total population sizes}}
\label{fig:SSD_absabu}
\end{figure*}

\begin{figure*}
\includegraphics[width=\textwidth]{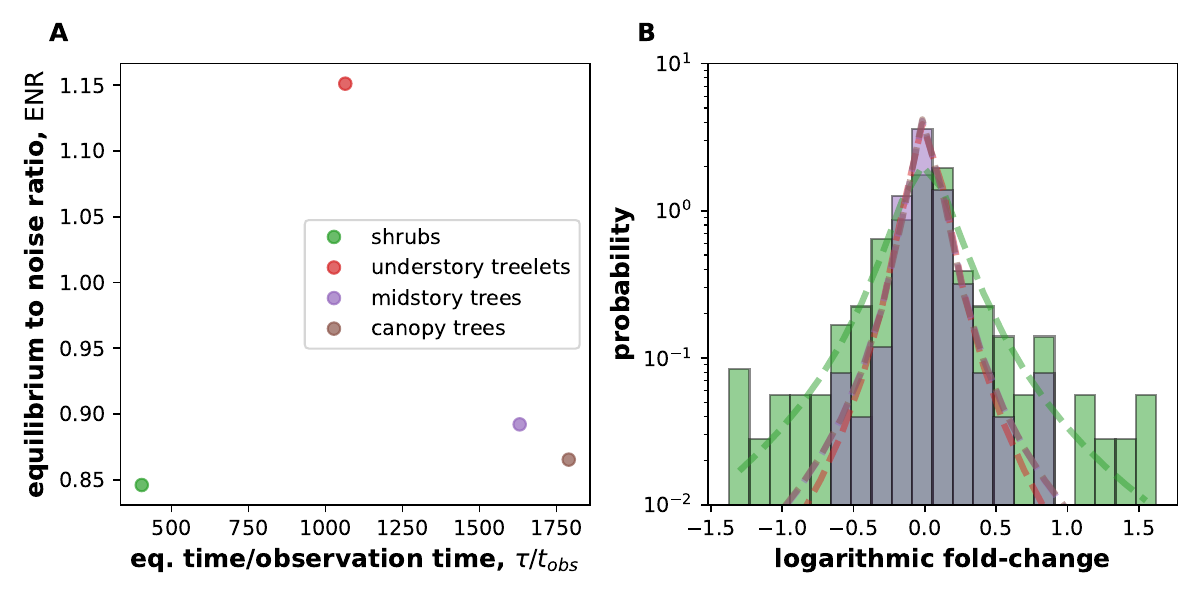}
\caption{\RV{\textbf{Comparing forest clusters.} \textbf{A)} The ENR and equilibration timescale $\tau$ obtained by fitting the LFD of the four tree clusters. Shrubs have a much shorter equilibration timescale than the other three clusters. The ENR varies over only a small range of values. \textbf{B)} The predicted LFD distributions at the best-fit parameters are plotted, along with the underlying data (only two of the four  clusters shown for clarity). The LFD of shrubs differs significantly from the LFDs of understory treelets, midstory trees, and canopy trees, which are all similar to each other. Notably, shrubs experience many more large fluctuations than the other three categories.} }
\label{fig:comparing_forest_clusters}
\end{figure*}

\FloatBarrier
\bibliography{bib_cities,refs}
\bibliographystyle{naturemag}